\documentclass[5p,times]{elsarticle}

\usepackage[numbers]{natbib}
\usepackage[nolist]{acronym}

\usepackage{fontawesome}
\usepackage{float}
\usepackage{placeins}
\usepackage{hyperref}
\usepackage{multirow}
\usepackage{subcaption}
\usepackage{soul}
\usepackage{color,soul}
\usepackage{graphicx}
\usepackage{tikz}
\usepackage{colortbl} 
\usepackage{threeparttable}
\usepackage{stfloats}
\usepackage{float}
\hypersetup{
    colorlinks=false,
    hidelinks
}

\usepackage{amssymb}
\usepackage{textcomp}
\usepackage{stfloats}
\usepackage{url}
\usepackage{verbatim}
\usepackage{graphicx}
\usepackage{acronym}
\usepackage{fontawesome}
\usepackage{float}
\usepackage{placeins}
\usepackage{amsmath}
\usepackage{multirow}
\usepackage{hyperref}
\hypersetup{
    colorlinks=false,
    linkcolor=black,
    citecolor=black,
    urlcolor=black
}
\usepackage{graphicx}
\usepackage{tikz}
\newcommand{\circlednumber}[1]{%
    \tikz[baseline=(char.base)]{
        \node[shape=circle, draw=black, fill=black, text=white, inner sep=0.6pt, font=\sffamily\small] (char) {#1};
    }
}
\usepackage[utf8]{inputenc}
\usepackage[T1]{fontenc}
\usepackage[numbers]{natbib}
\usepackage{fontawesome}
\usepackage[english]{babel}

\journal{Future Generation Computer Systems}

\begin{document}
\acrodef{AI}{Artificial Intelligence}
\acrodef{API}{Application Programming Interface}
\acrodef{APIs}{Application Programming Interfaces}
\acrodef{ANOVA}{Analysis of Variance}
\acrodef{CERN}{European Organization for Nuclear Research}
\acrodef{CPU}{Central Processing Unit}
\acrodef{CNN}{Convolutional Neural Network}
\acrodef{CNNs}{Convolutional Neural Networks}
\acrodef{DNN}{Deep Neural Network}
\acrodef{DNNs}{Deep Neural Networks}
\acrodef{DRL}{Deep Reinforcement Learning}
\acrodef{DT}{Decision Tree}
\acrodef{FIBRE-NG}{Future Internet Brazilian Environment for Experimentation New Generation}
\acrodef{GNN}{Graph Neural Networks}
\acrodef{GPN}{Great Plains Network}
\acrodef{I/O}{Input/Output}
\acrodef{KNN}{K-Nearest Neighbors}
\acrodef{LSTM}{Long Short-Term Memory}
\acrodef{MAE}{Mean Absolute Error}
\acrodef{MLaaS}{Machine Learning as a Service} 
\acrodef{ML}{Machine Learning}
\acrodef{MOS}{Mean Opinion Score}
\acrodef{MAPE}{Mean Absolute Percentage Error}
\acrodef{MSE}{Mean Squared Error}
\acrodef{NMAE}{Normalized Mean Absolute Error}
\acrodef{QoE}{Quality of experience}
\acrodef{QoS}{Quality of Service}
\acrodef{RAM}{Random-Access Memory}
\acrodef{RF}{Random Forest}
\acrodef{RL}{Reinforcement Learning}
\acrodef{RMSE}{Root Mean Square Error}
\acrodef{SFI2}{Slicing Future Internet Infrastructures}
\acrodef{SDN}{Software-Defined Networking}
\acrodef{SLA}{Service-Level Agreement}
\acrodef{SL}{Supervised Learning}
\acrodef{TPE}{Tree-Structured Parzen Estimator}
\acrodef{UL}{Unsupervised Learning}
\acrodef{VoD}{Video on Demand}
\acrodef{WMAPE}{Weighted Mean Absolute Percentage Error}
\let\WriteBookmarks\relax
\def\floatpagepagefraction{1}
\def\textpagefraction{.001}
\begin{frontmatter}



\title{AI-driven Orchestration at Scale: Estimating Service Metrics on National-Wide Testbeds}


\author[1]{Rodrigo Moreira\corref{cor1}}
\ead{rodrigo@ufv.br}                        

\affiliation[1]{organization={Federal University of Viçosa (UFV)},
                city={Rio Paranaíba},
                state={Minas Gerais},
                country={Brazil}}

\cortext[cor1]{Corresponding author}    

\author[2]{{Rafael} Pasquini}

\ead{rafael.pasquini@ufu.br}

\affiliation[2]{organization={Federal University of Uberlândia (UFU)},
                city={Uberlândia},
                state={Uberlândia},
                country={Brazil}}

\author[3]{Joberto S. B. Martins}

\ead{joberto.martins@animaeducacao.com.br}

\affiliation[3]{organization={Salvador University (UNIFACS)},
                city={Salvador},
                state={Bahia},
                country={Brazil}}

\author[4]{Tereza C. Carvalho}

\ead{terezacarvalho@usp.br}

\affiliation[4]{organization={University of São Paulo (USP)},
                city={São Paulo},
                state={São Paulo},
                country={Brazil}}

\author[5]{Flávio de Oliveira Silva}

\ead{flavio@di.uminho.pt}



\affiliation[5]{organization={University of Minho (UMinho)},
                city={Braga},
                country={Portugal}}

\begin{abstract}
 Network Slicing (NS) realization requires AI-native orchestration architectures to efficiently and intelligently handle heterogeneous user requirements. To achieve this, network slicing is evolving towards a more user-centric digital transformation, focusing on architectures that incorporate native intelligence to enable self-managed connectivity in an integrated and isolated manner. However, these initiatives face the challenge of validating their results in production environments, particularly those utilizing ML-enabled orchestration, as they are often tested in \textcolor{black}{local networks or laboratory simulations}. This paper proposes a \textcolor{black}{large-scale validation method} using a network slicing prediction model to forecast latency using Deep Neural Networks (DNNs) and \textcolor{black}{basic ML algorithms} embedded within an NS architecture evaluated in real large-scale production testbeds. It measures and compares the performance of different DNNs and ML algorithms, considering a distributed database application deployed as a network slice over two large-scale production testbeds. The investigation highlights how AI-based prediction models can enhance network slicing orchestration architectures and presents a seamless, production-ready validation method as an alternative to \textcolor{black}{fully controlled simulations or laboratory setups.}
\end{abstract}








\begin{highlights}
\item Forecasting behavior in production-ready network slicing architectures.
\item Architectural study of embedded DNNs and basic ML for slicing SLA conformance;
\item Evaluation of hyperparameter tuning for AI-native network slices on testbeds;
\item Creation of a dataset workflow for realistic slicing application workloads;
\end{highlights}

\begin{keyword}
Network Slicing \sep Deep Neural Networks \sep Machine Learning \sep Service-Level Agreement \sep Distributed Database
\end{keyword}

\end{frontmatter}



\section{Introduction}\label{sec:introduction}

Modern applications require challenging behaviors from phys\-i\-cal networks to satisfy stringent requirements such as ultra-reliability, low latency, and high throughput~\cite{Liu2023}. In addition to these quantifiable network requirements, it is necessary to incorporate seamless, intelligent, and pervasive network capabilities to satisfy user demands ~\cite{Zhang2023, Li2023}. Although network management, control planes, and data planes have evolved to address this issue, challenges remain and require further large-scale evaluation.

Many approaches, technologies, and methods have been developed to build user-oriented network architectures that provide connectivity in an isolated and personalized manner~\cite{Jawad2023}. One key technological enabler of this vision is network slicing, which establishes network connectivity on top of physical infrastructure while ensuring isolation, end-to-end connectivity, and application-driven requirements, with \textcolor{black}{dedicated} control and data planes~\cite{Moreira2021}. With this service-tailoring capability, \ac{ML} effectively addresses various management and orchestration challenges, thereby enabling intelligent and real-time insights \textcolor{black}{for service provider managers}. A major advantage of intelligent network orchestration is the ability of \ac{AI} to ensure and evaluate the service quality and user experience in network slicing~\cite{Saibharath2023}.

\ac{AI} techniques, such as reinforcement learning, supervised learning, and unsupervised learning, have been effectively integrated with network orchestrators to mitigate cybersecurity threats, enable intelligent resource allocation, and ensure \ac{SLA} assurance for network slicing~\cite{MOREIRA2023108852, Ahmed2023, Haque2023, moreira2023b}. In network slicing \ac{SLA} assurance \textcolor{black}{and measurement}, computational intelligence techniques have been progressively incorporated into the building blocks of orchestration architectures. However, these architectures are neither natively secure nor inherently intelligent, often leading to context-specific solutions~\cite{Wu2022}. \textcolor{black}{Consequently, they remain dependent on third-party vendors, raising concerns regarding the security and privacy of network slicing services.} 

\textcolor{black}{A key challenge in incorporating intelligence into network slicing architectures for the effective utilization of \ac{AI} pipelines is ensuring data quality and granularity~\cite{SINGH2023144} while preserving user privacy and avoiding the exposure of network slicing application details. Our approach addresses this by leveraging generic infrastructure metrics \textcolor{black}{appropriately combined} to enable \textcolor{black}{\ac{SLA} fitting and forecasting} across diverse network slicing applications. Although demonstrated through \ac{SLA} prediction for Cassandra applications, the proposed methodology is designed with principles and techniques that can be generalized to various slicing applications.} Evaluating the effectiveness of this intelligent orchestration mechanism requires analyzing network environments under conditions comparable to those of a production-ready network~\cite{Gomez2023}.


This paper investigated the performance of our intelligent native orchestration architecture in production-ready network scenarios under realistic conditions. Hence, we developed and \textcolor{black}{enhanced} an orchestration framework for network slicing~\cite{Martins2023}. We evaluated the capability of the newer \texttt{Predictor Module} within the ``\ac{SFI2} AI Management'' building block to forecast latency from a \textcolor{black}{distributed database (Cassandra) deployed as a network slice}, operating on nationwide testbeds FIBRE-NG~\cite{salmito2014fibre} and Fabric~\cite{Baldin2019}. \ac{DNNs} and basic \ac{ML} algorithms were employed to predict Cassandra latency within our dataset generation framework.

This paper presents several key contributions: (1) an empirical analysis of \ac{DNNs} \textcolor{black}{and basic ML algorithms} for network slicing latency forecasting on large-scale testbeds, (2) an in-depth investigation of hyperparameter tuning for \ac{DNNs}, (3) the development of a realistic network application dataset with a comprehensive workload, and (4) a detailed deployment workflow for network slicing services on nationwide testbeds.

The structure of the remainder of this paper is as follows. In Section~\ref{sec:related_work}, we contextualize our work within a broader scope, highlighting the unique contributions of this research. The proposed method is presented in detail in Section~\ref{sec:proposed_method}, followed by a description of the experimental setup in Section~\ref{sec:experiments_and_model_computation}. Section~\ref{sec:evaluation_results} discusses the results, and  Section~\ref{sec:concluding_remarks} presents the concluding remarks and future directions.

\section{Related Work}\label{sec:related_work}

In the literature, there are different uses of \ac{ML} for predicting \ac{QoE} or \ac{QoS}, both for applications in the context of mobile networks and generic applications on network slices~\cite{Phyu_2023}. Some approaches use machine learning algorithms pre-trained in simulation environments, validate how these algorithms behave, and compute performance metrics~\cite{Moreira_2023}. Other approaches have been proposed for network simulation environments, leaving aside the validation of these methods in real network testbeds \cite{Donatti_2023}. 

Pasquini and Stadler~\cite{pasquini_learning_2017} combined three metrics to estimate application \ac{QoE} while running different applications in an OpenFlow network. They combined the operating system metrics, network flows in switch tables, and application metrics to train machine learning. The experiments considered two applications: \ac{VoD} and Voldemort-distributed databases. They simulated two traffic behaviors to assess the suitability of QoE prediction by considering the observed metrics. Our approach further explores and validates whether \ac{ML} behaves properly in a real network testbed.

In network slicing orchestration matter, Cui \textit{et} al.~\cite{cui_qos_2022} proposed reinforcement learning combined with the \ac{LSTM} algorithm to guarantee \ac{QoS} for vehicle-to-vehicle network slicing. The search space of \ac{DRL} is a 5G antenna, where the action relies on the number of resource blocks. Our approach estimates \ac{QoE} in a real network testbed.

Nougnanke \textit{et} al.~\cite{nougnanke_ml-based_2023} proposed and evaluated an \ac{ML}-based approach for modeling and predicting network traffic under different workloads. They can estimate latency under different network conditions using \ac{ML} algorithms. They used a mininet and a simulated environment to validate their findings using three different datasets. Our approach involves creating a dataset to validate the estimation of some network metrics; however, our approach focuses on estimating the \ac{QoE} in a real sliced network.

Ge \textit{et} al.~\cite{ge_forecasting_2023} proposed an \ac{GNN}-based end-to-end delay estimator for \ac{SDN} environments. They used a real dataset from the GEANT network to fit their model. In addition, as a proof of concept, they used the OMNet++ simulation environment to assess the prediction accuracy of the model for QoS assurance. A similar study using the same method as described above was differentiated using the Abline dataset \cite{ge_gnn-based_2022}.

Laiche \textit{et} al.~\cite{laiche_when_2021} introduced a multifactor influence in video \ac{QoE}. They used machine learning to predict  \ac{QoE} assurance in an emulated network scenario. They used and compared the performance of the \ac{KNN}, \ac{RF}, and \ac{DT} algorithms to predict user experience by considering popularity and user engagement on a well-known streaming web platform.

Abdelwahed \textit{et} al. ~\cite{abdelwahed_monitoring_2023} proposed and evaluated an ML-based approach for estimating Web \ac{QoE} using different context metrics. The metrics considered were network, browsing, and web user engagement. Using \ac{RF}, \ac{DT}, and \ac{KNN}, they estimated \ac{MOS} by considering user-side metrics. Similarly, our approach aims to predict the application \ac{QoE} by considering different metrics. 

Regarding \ac{QoS} estimation in network slicing using \ac{ML} algorithms, Khan \textit{et} al.~\cite{khan_highly_2022} applied and evaluated \ac{DNN} methods to select better network slicing according to connectivity services (mMTC, URLLC, and eMBB) requirements. Their experiments assessed how the \ac{ML} algorithm handled accurate slice assignment, slice-load balancing, and slice-failure scenarios. Instead of using a simulated network, our approach evaluates the effectiveness of \ ac{ML} in a real network to predict the applications \ac{QoE}.

For 5G mobile network slicing, Thantharate and Beard~\cite{thantharate_adaptive6g_2022} proposed and evaluated a transfer learning method for network slicing \ac{QoS}. Their proposed rationale is based on training local models for different slice network requirements in the source domain. The model can predict the resources in the target domain to satisfy slicing quality agreement. Our approach further predicts the application of \ac{SLA} conformance for an online network slice application in a real network testbed. Similarly, N. P. Tran \textit{et} al.~\cite{tran_ml_2023} proposed and evaluated an approach to estimate end-to-end throughput in 5G and B5G using \ac{LSTM}.


Yu \textit{et} al.~\cite{yu_network_2023} idealized a linear regression algorithm combined with \ac{RL} to predict slice mobility while ensuring \ac{QoS} for the application, minimizing costs, and maximizing revenues and profits. They used different statistics, such as user demand for CPU, RAM, and slicing resources to train the ML algorithm offline. Our approach focuses on predicting application \ac{QoE} through time-series regression, which enables multi-step predictions over time.

Yang \textit{et} al.~\cite{chiariotti_temporal_2023} bring a method for predicting \ac{QoS} for Virtual Reality applications in a network slicing using different machine learning algorithms. Their solutions can predict the latency, bandwidth, and different video codecs for different slicing models. Our approach aims to predict the Cassandra application \ac{SLA} in a real network testbed by using \ac{ML} algorithms.


\textcolor{black}{Dangi and Lalwani~\cite{Dangi2024} proposes a hybrid deep learning model for efficient network slicing in 5G networks. The model combines Harris Hawks Optimization (HHO) with Convolutional Neural Networks (CNN) and Long Short-Term Memory (LSTM) networks to optimize hyperparameters and classify network slices. The methodology involves three phases: loading the dataset, optimizing it with HHO, and classifying slices using the hybrid model. The results demonstrate that the proposed model outperforms existing methods in predicting appropriate network slices and offers improved service quality and efficiency in 5G network slicing.}

\textcolor{black}{Baktır \textit{et} al.~\cite{Baktır2024} proposes a Mixed Integer Programming (MIP) model and a heuristic algorithm called NESECS to optimize network slicing for various service types with different performance requirements. The MIP model aims to provide optimal solutions for small network instances, whereas NESECS is designed to handle larger instances efficiently. This study evaluates the performance of these solutions through extensive experiments, demonstrating that the proposed methods can effectively manage network resources, reduce SLA violations, and improve the overall network performance by ensuring logical isolation and resource reservation for different service types.}

\begin{table*}[htbp]
\caption{Related works and their properties.}
\label{tab:sota_summary}
\resizebox{\textwidth}{!}{%
\begin{tabular}{cccccc}
\hline
\textbf{Paper}                                          & \textbf{\begin{tabular}[c]{@{}c@{}}Experimental \\ Environment\end{tabular}} & \textbf{\begin{tabular}[c]{@{}c@{}}Evaluation \\ Metrics\end{tabular}}                                & \textbf{Dataset}                                                                      & \textbf{\begin{tabular}[c]{@{}c@{}}Enabling \\ Technologies\end{tabular}}            & \textbf{Applications}                                                                       \\ \hline
\cite{cui_qos_2022}                                                 & \begin{tabular}[c]{@{}c@{}}Simulated \\ Network\end{tabular}                 & \begin{tabular}[c]{@{}c@{}}Convergence Time and\\ Mobility Velocity\end{tabular}                      & Own                                                                                   & LSTM-DDPG                                                                            & \begin{tabular}[c]{@{}c@{}}V2V \\ Communication\end{tabular}                                \\ \hline
\cite{nougnanke_ml-based_2023}                                           & \begin{tabular}[c]{@{}c@{}}Emulated \\ Network\end{tabular}                  & \begin{tabular}[c]{@{}c@{}}\ac{NMAE}, and\\ Prediction Time\end{tabular}                                   & \begin{tabular}[c]{@{}c@{}}Telecom Italia \\ Big Data Challenge dataset~\cite{barlacchi_multi-source_2015} \end{tabular}                                                    & \begin{tabular}[c]{@{}c@{}}SDN, P4, NS3, \\ and ML Algorithms.\end{tabular}          & \begin{tabular}[c]{@{}c@{}}Network \\ Connectivity\end{tabular}                             \\ \hline
\cite{ge_forecasting_2023}                                                  & \begin{tabular}[c]{@{}c@{}}Emulated \\ Network\end{tabular}                  & \begin{tabular}[c]{@{}c@{}}\ac{RMSE}, \\ \ac{MAE}, \\ and Weighted \\ Mean Absolute Percentage Error (WMAPE).\end{tabular}                                      & GEANT                                                                                 & \begin{tabular}[c]{@{}c@{}}GNN, MLR, XGBoost, \\ and RF\end{tabular}                 & \begin{tabular}[c]{@{}c@{}}Network \\ Connectivity\end{tabular}                             \\ \hline
\cite{ge_gnn-based_2022}                                                  & \begin{tabular}[c]{@{}c@{}}Simulated \\ Network\end{tabular}                 & \begin{tabular}[c]{@{}c@{}}Packet Delay, \ac{MSE}, \\ \ac{RMSE}, \ac{MAE}, \\ and WMAPE.\end{tabular}                   & \begin{tabular}[c]{@{}c@{}}Abilene \\ Network\end{tabular}                            & \begin{tabular}[c]{@{}c@{}}GNN, \ac{RF}, \\ and NN\end{tabular}                           & \begin{tabular}[c]{@{}c@{}}Network \\ Connectivity\end{tabular}                             \\ \hline
\cite{pasquini_learning_2017}                                            & \begin{tabular}[c]{@{}c@{}}Local \\ Network\end{tabular}                     & \begin{tabular}[c]{@{}c@{}}\ac{NMAE}, and\\ Training Time\end{tabular}                                     & Own                                                                                   & \begin{tabular}[c]{@{}c@{}}OpenFlow,\\  and \ac{ML} Algorithms\end{tabular}               & \begin{tabular}[c]{@{}c@{}}\ac{VoD},\\  and KV\end{tabular}                                      \\ \hline
\cite{laiche_when_2021}                                              & \begin{tabular}[c]{@{}c@{}}Simulated \\ Network\end{tabular}                 & \begin{tabular}[c]{@{}c@{}}\ac{RMSE}, Correlation, \\ and Outlier Ratio\end{tabular}                       & Own                                                                                   & \begin{tabular}[c]{@{}c@{}}\ac{KNN}, \\ Decision Three (DT)\end{tabular}                  & VoD                                                                                         \\ \hline
\cite{abdelwahed_monitoring_2023}                                          & \begin{tabular}[c]{@{}c@{}}Local \\ Network\end{tabular}                     & \begin{tabular}[c]{@{}c@{}}Accuracy, Correlation,\\  and MOS rating.\end{tabular}                     & Own                                                                                   & \begin{tabular}[c]{@{}c@{}}KNN, RF, \\ and DT\end{tabular}                           & Web                                                                                         \\ \hline
\cite{khan_highly_2022}                                                & \begin{tabular}[c]{@{}c@{}}Simulated \\ Network\end{tabular}                 & \begin{tabular}[c]{@{}c@{}}Accuracy, Recall, Precision, \\ and F-1 score\end{tabular}                 & \begin{tabular}[c]{@{}c@{}}DeepSlice, \\ and Secure5G\end{tabular}                    & \begin{tabular}[c]{@{}c@{}}\ac{CNN}, \\ \ac{LSTM} and \\ Python-based ML frameworks.\end{tabular} & \begin{tabular}[c]{@{}c@{}}5G\\  Verticals\end{tabular}                                     \\ \hline
\cite{thantharate_adaptive6g_2022}                                         & \begin{tabular}[c]{@{}c@{}}Simulated \\ Network\end{tabular}                 & \begin{tabular}[c]{@{}c@{}}MSE, \\ Correlation\end{tabular}                                           & Own                                                                                   & \begin{tabular}[c]{@{}c@{}}Matlab Deep \\ Learning Toolbox.\end{tabular}             & \begin{tabular}[c]{@{}c@{}}5G \\ Verticals\end{tabular}                                     \\ \hline
\cite{yu_network_2023}                                                  & \begin{tabular}[c]{@{}c@{}}Simulated \\ Network\end{tabular}                 & \begin{tabular}[c]{@{}c@{}}RL-Convergence \\ Time\end{tabular}                                        & Own                                                                                   & NetworkX                                                                             & \begin{tabular}[c]{@{}c@{}}5G \\ Verticals\end{tabular}                                     \\ \hline
\cite{tran_ml_2023}                                                & \begin{tabular}[c]{@{}c@{}}Simulated \\ Network\end{tabular}                 & \begin{tabular}[c]{@{}c@{}}\ac{MAPE}, \\ packet loss, and delay\end{tabular}             & Own                                                                                   & \begin{tabular}[c]{@{}c@{}}\ac{LSTM}, \\ and Traditional ML Models\end{tabular}           & \begin{tabular}[c]{@{}c@{}}5G \\ Verticals\end{tabular}                                     \\ \hline
\cite{chiariotti_temporal_2023}                                          & \begin{tabular}[c]{@{}c@{}}Simulated \\ Network\end{tabular}                 & Accuracy                                                                                              & Own                                                                                   & \begin{tabular}[c]{@{}c@{}}Linear Regression \\ methods\end{tabular}                 & \begin{tabular}[c]{@{}c@{}}Virtual \\ Reality\end{tabular}                                  \\ \hline
\textcolor{black}{\cite{Dangi2024}}                                      & \textcolor{black}{\begin{tabular}[c]{@{}c@{}}Simulated \\ Network\end{tabular}}                 & \textcolor{black}{\begin{tabular}[c]{@{}c@{}}Accuracy, Precision, \\ Recall, and  F1-Score\end{tabular}}                 & \textcolor{black}{\begin{tabular}[c]{@{}c@{}}Unicauca IP Flow, \\ and 5G Network   Slicing\end{tabular}} & \textcolor{black}{\begin{tabular}[c]{@{}c@{}}HHO, CNN, \\ and LSTM\end{tabular}}                        & \textcolor{black}{\begin{tabular}[c]{@{}c@{}}Network \\ Slicing\end{tabular}}                                  \\ \hline
\textcolor{black}{\cite{Baktır2024}}                                              & \textcolor{black}{\begin{tabular}[c]{@{}c@{}}Simulated \\ Network\end{tabular}}                 & \textcolor{black}{\begin{tabular}[c]{@{}c@{}}SLA Violations, End-to-End Delay, \\ and Resource Utilization\end{tabular}} & \textcolor{black}{\begin{tabular}[c]{@{}c@{}}Custom \\ Simulated Data\end{tabular}}                      & \textcolor{black}{\begin{tabular}[c]{@{}c@{}}SDN, Virtualization, \\ and Edge Computing\end{tabular}}   & \textcolor{black}{\begin{tabular}[c]{@{}c@{}}Network Slicing, \\ and Edge Computing Optimization\end{tabular}} \\ \hline
\begin{tabular}[c]{@{}c@{}}\textbf{Our} \\ \textbf{Approach}\end{tabular} & \begin{tabular}[c]{@{}c@{}}\textbf{Production-ready}\\ \textbf{Network}\end{tabular}           & \begin{tabular}[c]{@{}c@{}}\textbf{\ac{MAE}}, \textbf{\ac{MAPE}}, \\ \textbf{and} \textbf{\ac{MSE}}\end{tabular}                                         & \textbf{Own}                                                                                   & \begin{tabular}[c]{@{}c@{}}\textbf{\ac{DNNs}}, \textbf{and}\\ \textbf{Basic ML Models}\end{tabular}                  & \begin{tabular}[c]{@{}c@{}}\textbf{Distributed}\\ \textbf{Database on a Sliced Testbed}\end{tabular}          \\ \hline
\end{tabular}%
}
\end{table*}

Approaches in the literature close to this proposal are summarized in Table~\ref{tab:sota_summary}. In this table, we have an ``Experimental Environment'' column related to which network environment the \ac{ML} algorithm makes \ac{QoS} or \ac{QoE} predictions. Many approaches have used simulated environments or local networks. The ``Evaluation Metrics'' column aims to show which metrics (\ac{MAPE}, \ac{MSE}, \ac{RMSE}, and others) the authors most often take into account to validate their proposals. The ``Dataset'' column indicates the dataset that each approach considers in the model training phase. Some approaches have created their own datasets, whereas others use datasets created by third parties. The ``Enabling Technologies'' column aims to identify which methods and technologies for both machine learning and simulating network environments are state-of-the-art. Many approaches still use classic ML approaches such as \ac{RF}, \ac{DT}, and \ac{KNN}. Finally, the ``Application'' column reports the applications for which the prediction engines estimate \ac{QoS} and \ac{QoE}. Some approaches focus on 5G verticals, while others focus on specific applications.

\section{Proposed Method}\label{sec:proposed_method}

Deploying network slices across multiple domains still requires advanced management and orchestration technologies capable of influencing the underlying network, particularly when dealing with heterogeneous devices. \textcolor{black}{Existing tools and techniques for implementing dynamic and elastic slices remain inadequately managed}, presenting opportunities for improvement, especially with \ac{AI} as a foundational enabler of such architectures. In this context, we previously proposed the Slicing Future Internet Infrastructures (\ac{SFI2}) reference architecture to manage and orchestrate \ac{AI}-native network slices while integrating diverse testbeds~\cite{Martins2023}.

\subsection{\ac{SFI2} Slicing Architecture}

\begin{figure}[htbp]
  \centering
  \includegraphics[width=0.99\columnwidth]{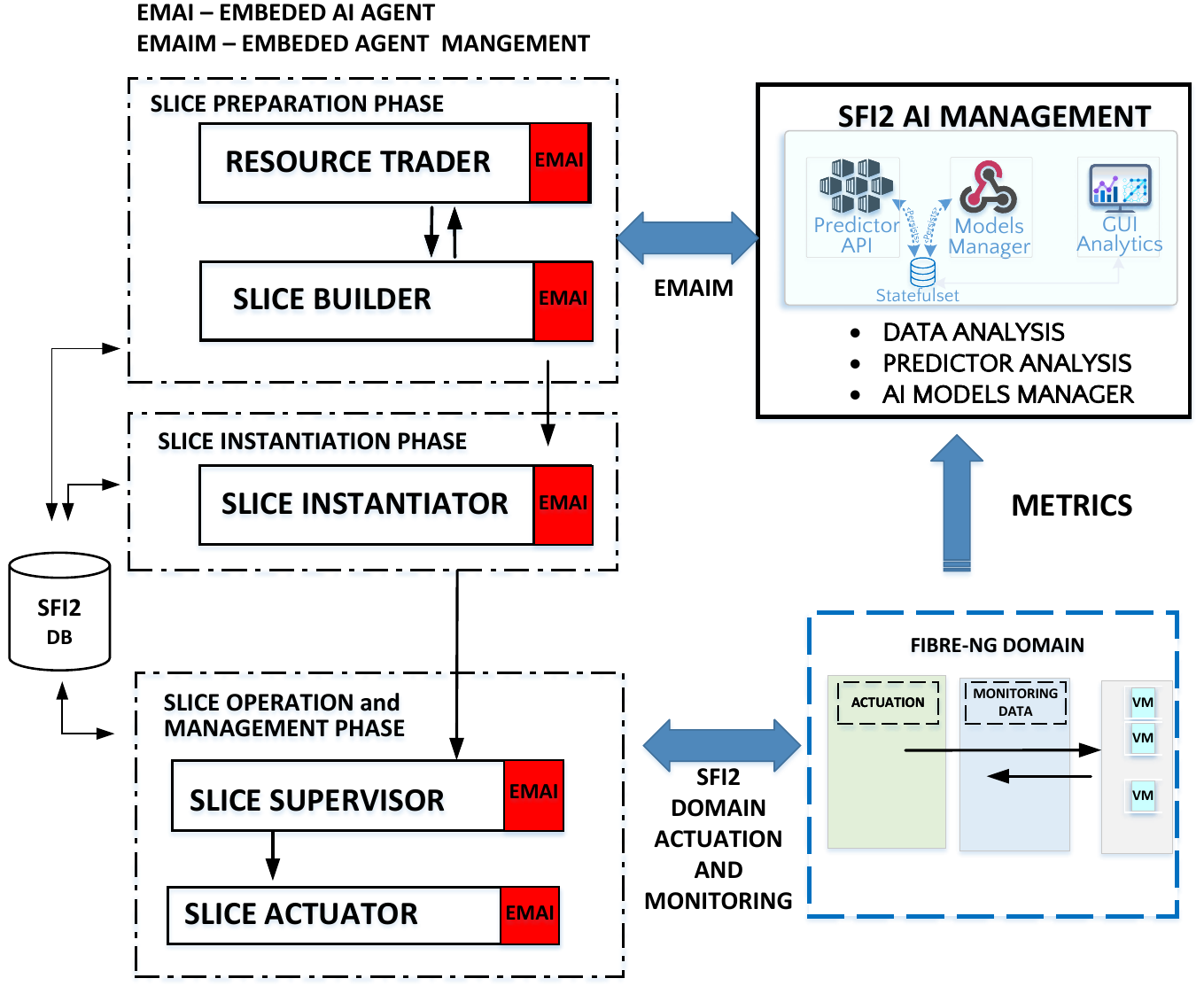}
  \caption{SFI2 Artificial Intelligence Agents Management within FIBRE-NG Domain Slices}
  \label{fig:SFI2Embed}
\end{figure}

The SFI2 architectural approach with native artificial intelligence embedded agents is illustrated in Fig. \ref{fig:SFI2Embed}. AI-native embedded agents in the architecture target the preparation, commissioning, operation, and decommissioning phases of the network slicing life-cycle.

Enhancements promoted by \ac{AI}-embedded agents include predicting resource and performance parameters using distinct \ac{AI} techniques, such as combinatorial optimization, \textcolor{black}{reinforcement learning, and neural networks}. Notably, a potential focus of \ac{AI}-embedded agents is the orchestration process involving different steps and actions applied to instantiate and dynamically reconfigure slices according to user requirements. 

Furthermore, the SFI2 architecture aims to operate over heterogeneous infrastructure following the concept of \ac{MLaaS}. To achieve this, \textcolor{black}{the SFI2 AI management module collects metrics from the target domains. It interacts with the embedded agents} in the functional blocks to manage the learning model and the other required parameters. The SFI2 AI management module manages learning agents throughout all infrastructure components to support the training, prediction, or decision tasks.

\subsection{Problem Setting and Method}\label{subsec:method_overview}

This paper proposes the evaluation of network slicing latency forecasting in a sliced SFI2-conform large-scale produc\-tion-ready testbed (FIBRE-NG and Fabric). The focus is on the  \ac{SFI2} \ac{AI} management functional block, which natively and \textcolor{black}{intelligently orchestrates slices} to estimate the \ac{SLA} compliance of an application running on a network slice.

\begin{figure*}[htb]
  \centering
  \includegraphics[width=0.8\textwidth]{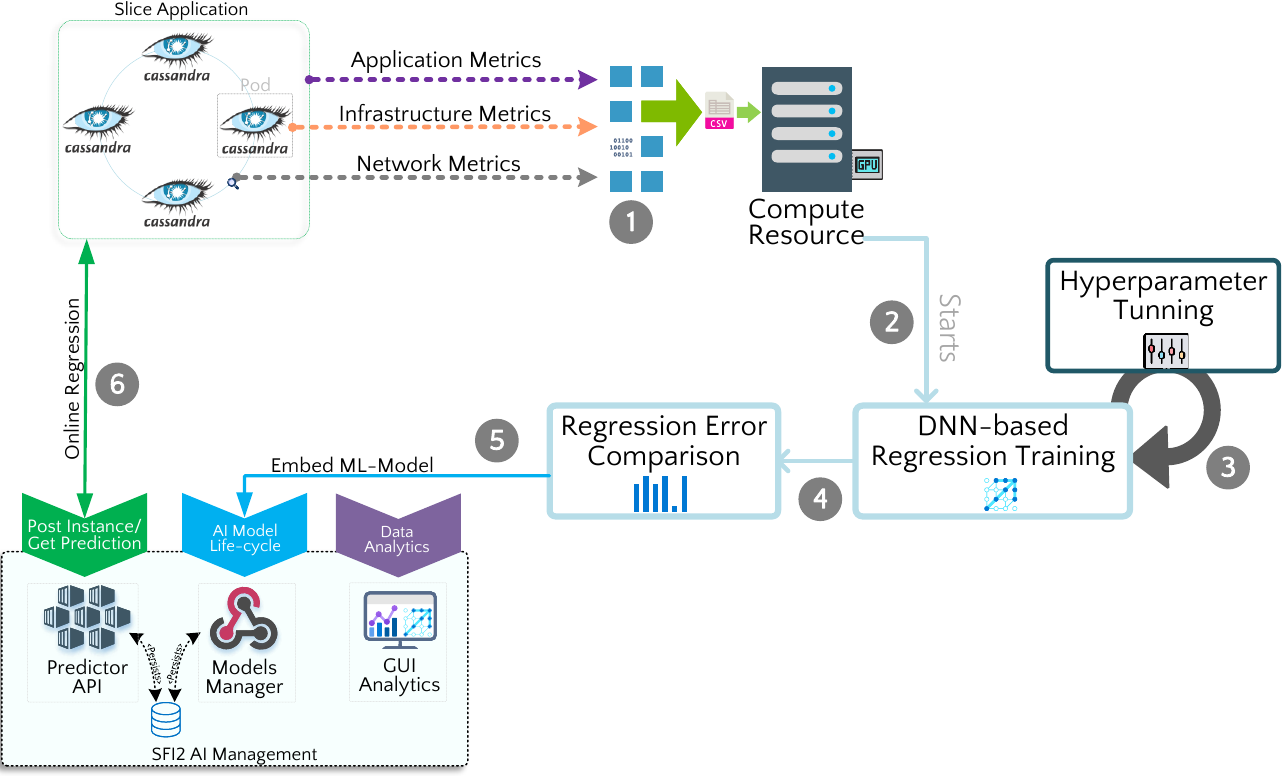}
  \caption{Proposed Method: Dataset Generation combined with \ac{DNNs} training and test flow.}
  \label{fig:method}
\end{figure*}

The proposed method is shown in Fig. ~\ref{fig:method}, highlighting the functional block of the orchestrator architecture and its interaction with the network slice lifecycle.


Thus, adopting the premise that the slice is implemented by the \ac{SFI2} Orchestrator \cite{Martins2023}, the procedures for online forecasting of \ac{QoE} begin by considering different metrics in different testbeds. The step \circlednumber{1} of the method refers to the collection and aggregation procedure of different $X$ metrics from the application, the network slice, and the underlying computational infrastructure. Application metrics $X_{application}$ refer to the statistics provided by applications at different execution stages. The $X_{cluster}$ infrastructure metrics refer to the computational resource consumption metrics demanded by the network slices from the underlying hardware. Network metrics $X_{network}$ refer to the network statistics of the network slice. These metrics are aggregated by data cleaning and standardization algorithms in the computational resource, resulting in a dataset $X = X_{application} \cup X_{cluster} \cup X_{network}$.

Cassandra, a distributed key-value database, is the application running on top of the network slicing used in this study. Using the \texttt{cassandra-stress}~\cite{cassandra_stress} tool, we generated logs of the reading and writing operations in this database to build a set of metrics $X_{application}$. \texttt{cassandra-stress} presents different performance metrics for reading and writing operations in the database, such as latency, operation rate, errors, and others.

The $X_{cluster}$ represents the statistics collected from the underlying infrastructure that supports the network slice execution. The metrics collected refer to the consumption of the \ac{CPU}, \ac{RAM}, and \ac{I/O} operations required by the network slice from the infrastructure. We also used the NetData monitoring framework to collect \ac{CPU}, \ac{RAM}, and \ac{I/O} metrics, as well as other metrics related to computational resources.

The $X_{network}$ variable refers to the \textcolor{black}{generic} statistics collected from the network interfaces of each \textcolor{black}{distributed} entity comprising the network slice. We employ a native network metrics extractor (NetData) to gather statistics on the volume of transmitted, received, and lost data across the various interfaces that support the Cassandra service. These individual network interface statistics are then aggregated based on their respective timestamps.

The $Y$ \textit{service-level metrics} for the Cassandra application are defined by the mean latency of the Read (R) and Write (W) operations. This latency represents the time taken by the Cassandra application to complete R and W operations when induced by \texttt{cassandra}-\texttt{stress}. During the training phase of \ac{DNNs} and \textcolor{black}{basic ML algorithms}, we extract features such as response time, errors, operations per second, and other relevant metrics from $X_{application}$ to construct the dataset. Conversely, in the $Y$ testing phase, our objective is to estimate the mean latency of various operations in Cassandra \textcolor{black}{by considering only the generic infrastructure metrics of the testbeds.}

Step \circlednumber{2} involves training using various algorithms based on \ac{ML} for time-series regression. We employ basic \ac{ML} and \ac{DNNs} to leverage their potential in handling complex datasets that lack linear relaxations, exhibit high-dimensional characteristics, and seamlessly adapt to new scenarios, thereby facilitating knowledge transfer.

Step \circlednumber{3} represents the search and adjustment of the hyperparameters for each \ac{DNN}. At this stage, the hyperopt~\cite{bergstra2023} tool uses Bayesian methods to find better parameters for training \ac{DNNs}, such as learning rate, batch size, and epochs. At the end of this phase, the models are trained with the best hyperparameters and exported to enable inference through the \ac{SFI2} \ac{AI} Management functional block.

The \circlednumber{4} step involves the empirical comparison of \textcolor{black}{basic \ac{ML} algorithms} and \ac{DNNs} using appropriate metrics for regression problems, such as \ac{MAE}, \ac{MSE}, and \ac{MAPE}. \ac{MAE} represents the average of the absolute differences between the predicted and actual values, while \ac{MSE} is the mean of the squares of these differences, thereby emphasizing larger deviations in evaluating regression models. \ac{MAPE} is a measure of relative error that expresses the difference between actual values and those predicted by a regression model as a percentage. \ac{MAPE} is independent of the data scale, making it suitable for comparing the accuracy of regression models.

In step \circlednumber{5}, we train the models and fine-tune the hyperparameters before integrating the trained models into \ac{SFI2} \ac{AI} Management. The \ac{SFI2} architecture receives these models and makes them available for future \ac{SLA} forecasting using the Predictor API.

In step \circlednumber{6}, the \ac{API} of the \ac{SFI2} AI Management block can handle some instances of $X_{application, cluster, network}$ and return a possible condition of the \ac{QoE} $Y$ in which the network slice is conditioned online, allowing us to evaluate whether the \ac{SLA} is being honored.

\subsection{Service Infrastructure Statistics and Service-Level Mertics}\label{subsec:sevice_level_statistics}

In this section, we detail the set of input features $X =$  $X_{application}$ $ \cup X_{cluster}$ $ \cup X_{network}$ and denote the response variable $Y$. The $X_{application}$ statistics refer to the volume of operations performed on the database at a given timestamp, errors, and lines written per second for the Cassandra application. We extracted these metrics from the \textcolor{black}{\texttt{cassandra-stress} utility, thus linking each record of these statistics to the corresponding timestamp}. The $X_{cluster}$ statistics are metrics related to the consumption of the computational resources of each node that hosts the Cassandra application containers. These statistics are \textcolor{black}{\ac{CPU} consumption, \ac{RAM} memory, and interrupts of the host machines collected through NetData}.

The $X_{network}$ metric refers to the consumption of network resources by each computing node and container in the Cassandra application. Specifically, network statistics refer to the volume of data sent, errors received from the containers that run the Cassandra ring components, and computing nodes that host the application. These statistics were collected every second \textcolor{black}{using NetData}.

The $Y$ metric refers to the \ac{QoE} experienced by a user when operating on a distributed database. This metric considers the average latency of write (W) and read (R) operations in Cassandra deployed on the testbed. The response variable metric $Y$ was measured using the Cassandra-stress utility, and we linked each latency per second to a given timestamp. \textcolor{black}{All write and read operations of our dataset generation framework in distributed testbeds did not take into account application caches in any of the replicas.}

We model the collection of these metrics as $X$ and $Y$ time series so that our objective is to estimate the average of the W and R operations in the Cassandra application deployed by the \ac{SFI2} Architecture over different testbeds using a regression problem in supervised learning~\cite{James2013}. Then, we estimate $Y_{t}$ using machine learning algorithms that learn from the statistics $X_{t} = [X_{1t}, \cdots, X_{dt}]$. Thus, we find a model $M : X_{t} \rightarrow \hat{Y}_{t}$, where $\hat{Y}_{t}$ optimally approximates $Y_{t}$ for a given $X_{t}$.

We used different \textcolor{black}{\ac{ML} models and} \ac{DNNs} model architectures to solve the regression problem. \textcolor{black}{We used \ac{DT}, \ac{RF}, \ac{KNN}, \ac{LSTM}, CatBoost, and XGBoost.} Among them we use \ac{DNNs} such as: FCN~\cite{fcn}, FCNPlus~\cite{tsai}, ResNet~\cite{tsai}, ResNetPlus~\cite{tsai}, ResCNN~\cite{tsai}, TCN~\cite{tcn}, InceptionTime~\cite{inceptionTime}, InceptionTimePlus~\cite{inceptionTime}, OmniScaleCNN~\cite{Marc2020}, XCM~\cite{tsai}, and XCMPlus~\cite{tsai} that are implemented in the framework Fastai and \textit{tsai}~\cite{tsai, Howard2020}.

\textcolor{black}{Table~\ref{tab:dnn_structure_details} provides an overview of the employed \ac{DNNs} structure. The Trainable Parameters column indicates the number of parameters that can be adjusted during training, including the weights and biases. The total number of Layers refers to the overall number of layers in the model, encompassing convolutional, pooling, and fully connected layers. The kernel Sizes describe the dimensions of the filters used in the convolutional layers. Pooling highlights the presence of pooling layers, which help reduce data dimensionality. Mult-Adds (M) represents the computational cost measured in millions of multiplication and addition operations. Finally, the Estimated Size (MB) estimates the model’s size in megabytes, reflecting the number of parameters and required storage.}

\begin{table}[H]
\caption{\textcolor{black}{Detailed Structure of Employed DNN Models}}
\label{tab:dnn_structure_details}
\resizebox{\columnwidth}{!}{%
\begin{tabular}{|c|c|c|c|c|c|}
\hline
\textbf{Model}    & \textbf{\begin{tabular}[c]{@{}c@{}}Trainable \\ Parameters\end{tabular}} & \textbf{\begin{tabular}[c]{@{}c@{}}Total\\ Layers\end{tabular}} & \textbf{\begin{tabular}[c]{@{}c@{}}Kernel \\ Sizes\end{tabular}} & \textbf{\begin{tabular}[c]{@{}c@{}}Mult-\\ Adds (M)\end{tabular}} & \textbf{\begin{tabular}[c]{@{}c@{}}Estimated \\ Size (MB)\end{tabular}} \\ \hline
FCN               & 285,446                                                                  & 17                                                              & (7), (5), (3)                                                    & 14.47                                                             & 2.18                                                                    \\
FCNPlus           & 285,446                                                                  & 18                                                              & (7), (5), (3)                                                    & 14.47                                                             & 2.18                                                                    \\
ResNet            & 490,758                                                                  & 53                                                              & (7), (5), (3), (1) $\times$ 3                                    & 24.86                                                             & 3.74                                                                    \\
ResNetPlus        & 490,758                                                                  & 56                                                              & (7), (5), (3), (1) $\times$ 3                                    & 24.86                                                             & 3.74                                                                    \\
ResCNN            & 268,551                                                                  & 33                                                              & (7), (5), (3), (1), (3) $\times$ 3                               & 13.58                                                             & 2.05                                                                    \\
TCN               & 71,406                                                                   & 95                                                              & (7) $\times$ 2, (1), (7) $\times$ 13                             & 1.88                                                              & 0.54                                                                    \\
InceptionTime     & 460,038                                                                  & 90                                                              & (1), (39), (19), (9) $\times$ 6                                  & 23.32                                                             & 3.51                                                                    \\
InceptionTimePlus & 460,038                                                                  & 124                                                             & (1), (39), (19), (9) $\times$ 6                                  & 23.32                                                             & 3.51                                                                    \\
OmniScaleCNN      & 5,239,596                                                                & 68                                                              & (1), (2), (3), (5), (7), (11) $\times$ 3, (1), (2)               & 266.26                                                            & 39.97                                                                   \\
XCM               & 328,584                                                                  & 29                                                              & (51), (1), (51)                                                  & 24.64                                                             & 2.51                                                                    \\
XCMPlus           & 328,584                                                                  & 30                                                              & (51), (1), (51)                                                  & 24.64                                                             & 2.51                                                                    \\ \hline
\end{tabular}%
}
\end{table}

 The choice of these neural networks aimed to fulfill our objective of empirically comparing the performance of \ac{DNNs} in estimating the \ac{SLA} compliance. We also used the Optuna~\cite{Akiba2019} framework to optimize the hyperparameters of the \ac{DNNs} for comparison and employed the \ac{TPE}-based algorithm~\cite{Yang2020}. Our optimizer sought to find the optimal parameters according to the search space, as shown in Table~\ref{tab:search_space}. 

\begin{table}[H]
\centering
\caption{Optuna Search Space for each Regression Model (\ac{DNNs}).}
\label{tab:search_space}
\begin{tabular}{lc}
\hline
\rowcolor[HTML]{EFEFEF} 
\multicolumn{1}{c}{\cellcolor[HTML]{EFEFEF}\textbf{Hiperparameter}} & \textbf{\begin{tabular}[c]{@{}c@{}}Search\\ Space\end{tabular}} \\ \hline
\textbf{Batch Size}                                                 & 8, 16, 32                                                       \\
\rowcolor[HTML]{EFEFEF} 
\textbf{Learning Rate (LR)}                                         & 0.1, 0.01, 0.001                                                \\
\textbf{Epochs}                                                     & 20, 50, 100                                                     \\
\rowcolor[HTML]{EFEFEF} 
\textbf{Patience}                                                   & 5, 10, 50                                                       \\
\textbf{Optimizer}                                                  & Adam, SGD                                                       \\
\rowcolor[HTML]{EFEFEF} 
\textbf{\# of Layers}                                                & 1, 2, 3, 4, 5                                                   \\
\textbf{Hidden Size}                                                & 50, 100, 200                                                    \\
\rowcolor[HTML]{EFEFEF} 
\textbf{Bidirectional}                                              & \textit{True}, \textit{False}                                                     \\ \hline
\end{tabular}
\end{table}

\subsection{Dataset Generation}\label{subsec:dataset_generation}

To understand how \ac{ML} algorithms perform in real testbeds, we propose a dataset that generates workloads using a periodic-load pattern and a collection framework. Using \texttt{cassandra}-\texttt{stress}, we generated \textcolor{black}{\textbf{W}rite and \textbf{R}ead requests for the application deployed on the \ac{FIBRE-NG} and Fabric testbeds}. These requests follow a Poisson process where the request rate adheres to a sinusoidal function, starting with a level $P_{s}$ and amplitude $P_{A}$ until \textcolor{black}{500k lines are written or read from the Cassandra application}. \textcolor{black}{The initial Cassandra parameters were defined as follows: consistency level set to \textit{quorum}, replica factor of 2, and 256 tokens}.

\begin{figure}[htb]
  \centering
  \includegraphics[width=1\columnwidth]{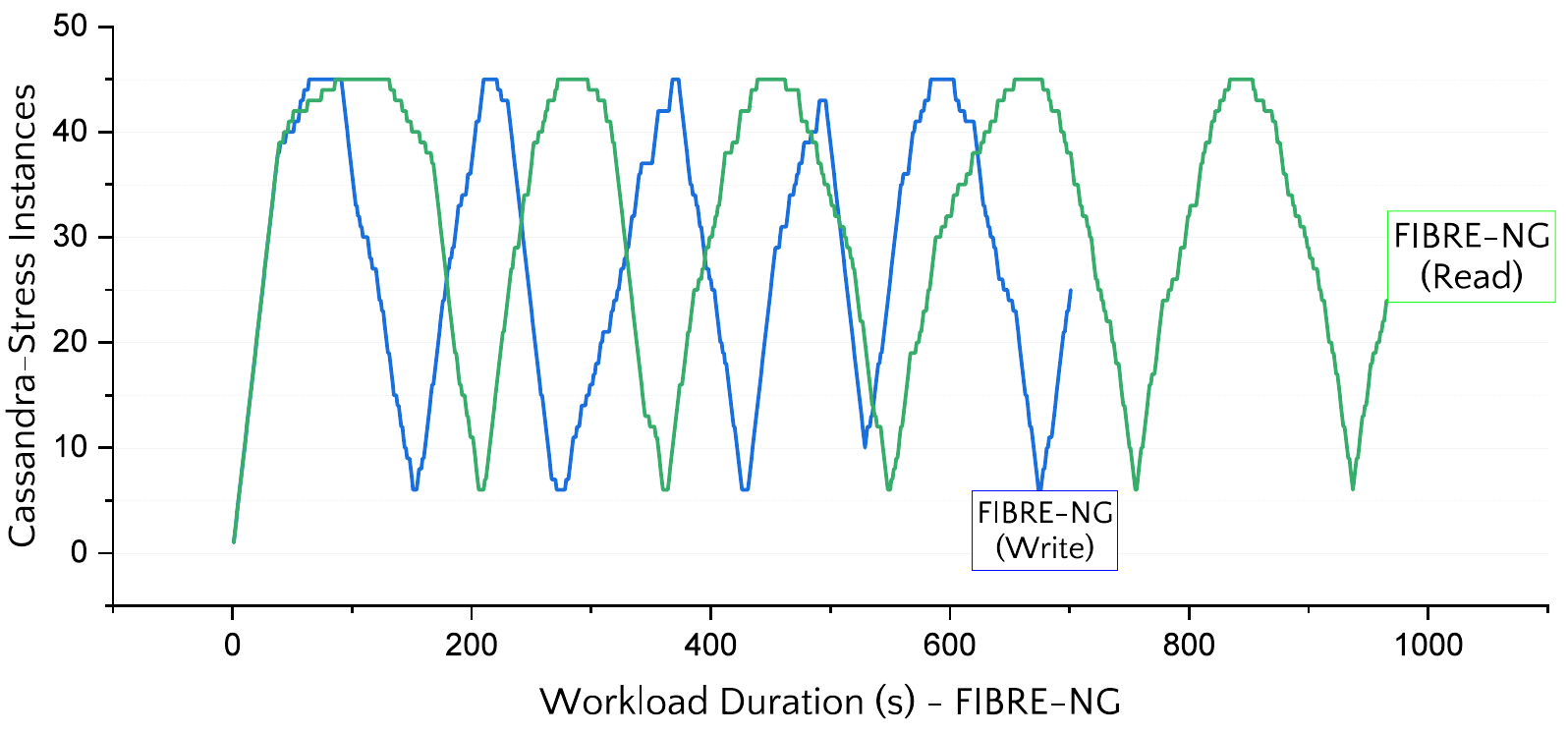}
  \caption{\textcolor{black}{FIBRE-NG Testbed -- Generating the traces.}}
  \label{fig:fibre_NG-generating_the_traces}
\end{figure}

\begin{figure}[htb]
  \centering
  \includegraphics[width=1\columnwidth]{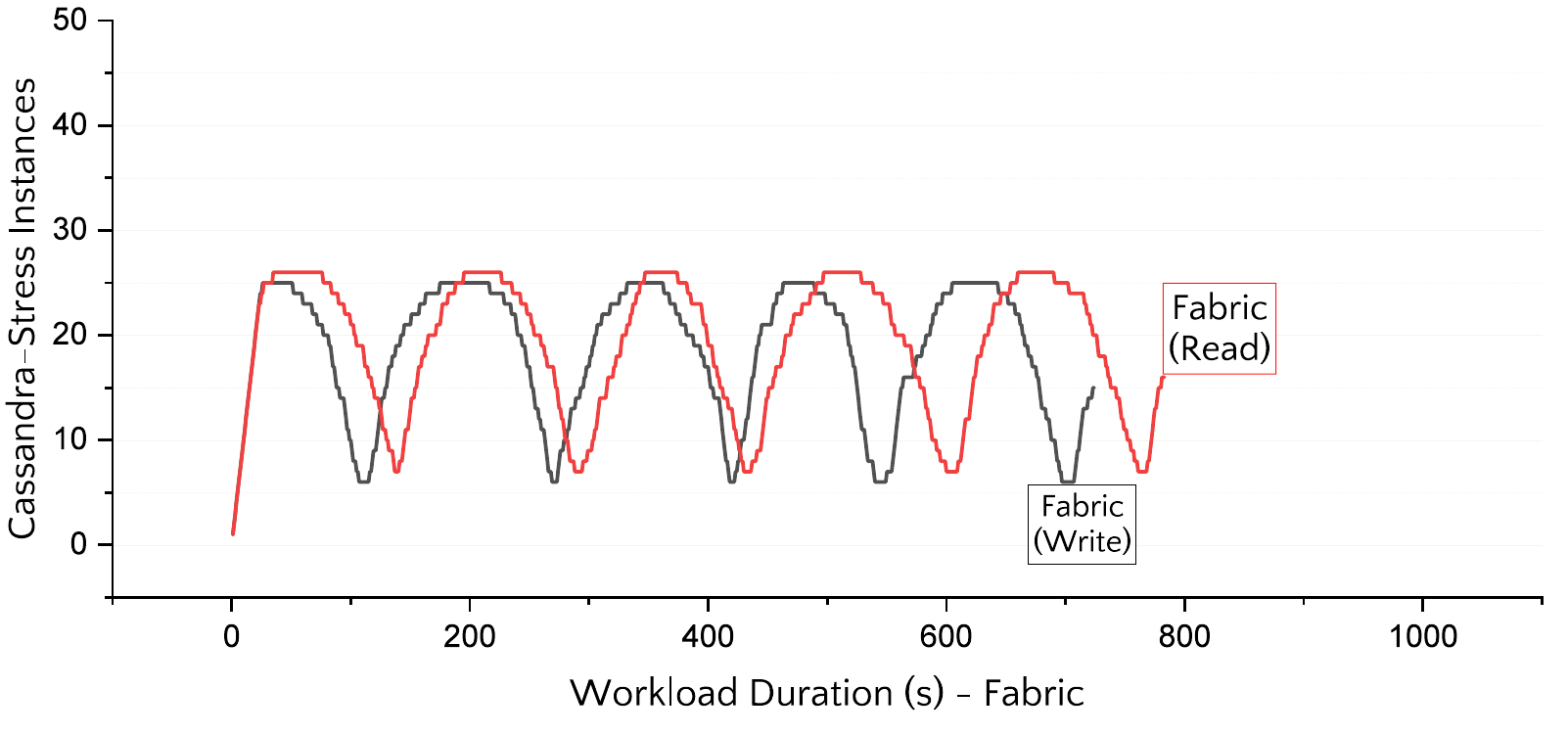}
  \caption{\textcolor{black}{Fabric Testbed -- Generating the traces.}}
  \label{fig:fabric-generating_the_traces}
\end{figure}

In Fig.~\ref{fig:fibre_NG-generating_the_traces}, and \ref{fig:fabric-generating_the_traces}, the workload pattern represents the number of \texttt{cassandra-stress} processes created according to a sinusoidal function over simulation time. The traces in Fig.~\ref{fig:fibre_NG-generating_the_traces} and \ref{fig:fabric-generating_the_traces} refer to the instances of \texttt{cassandra-stress}, generating requests of different types, such as \textbf{W}rite and \textbf{R}ead, to create application metrics. Whenever a new \texttt{cassandra-stress} process is created, it triggers requests (W or R) to the Cassandra application running on the testbeds. We empirically define the parameters of the sinusoidal function as $f(t) = 22.5 + \frac{45}{2} \sin\left(\frac{2\pi t}{T}\right)$, where \(t\) represents time and \(T\) is the period of the function.

This function models a wave that oscillates between 0 and 45, with an amplitude (\(P_A\)) of \(\frac{45}{2}\) units and an average value (\(P_s\)) of 22.5, providing an adequate representation for the desired variability in the generated processes. These values were defined empirically \textcolor{black}{because of the restriction of computational resources for \texttt{cassandra-stress} (container), and high values of $P_{s}$ and $P_{A}$ imply high resource consumption and can lead to container failure, thereby damaging the creation of the dataset}. Both Write and Read operations were performed in Cassandra after the warm-up process.

\textcolor{black}{Despite the Poisson model has limitations, particularly in representing peak load conditions, as it focuses on typical traffic rather than maximum loads. The dataset was generated in a live production environment using a synthetic workload application. The models were trained with realistic live production data, replicating actual network behaviors. This environment, with interconnected distributed nodes on a large-scale testbed, could experiences unexpected traffic patterns and anomalies.} 

\begin{figure}[htbp]
    \centering
    \includegraphics[width=0.9\columnwidth]{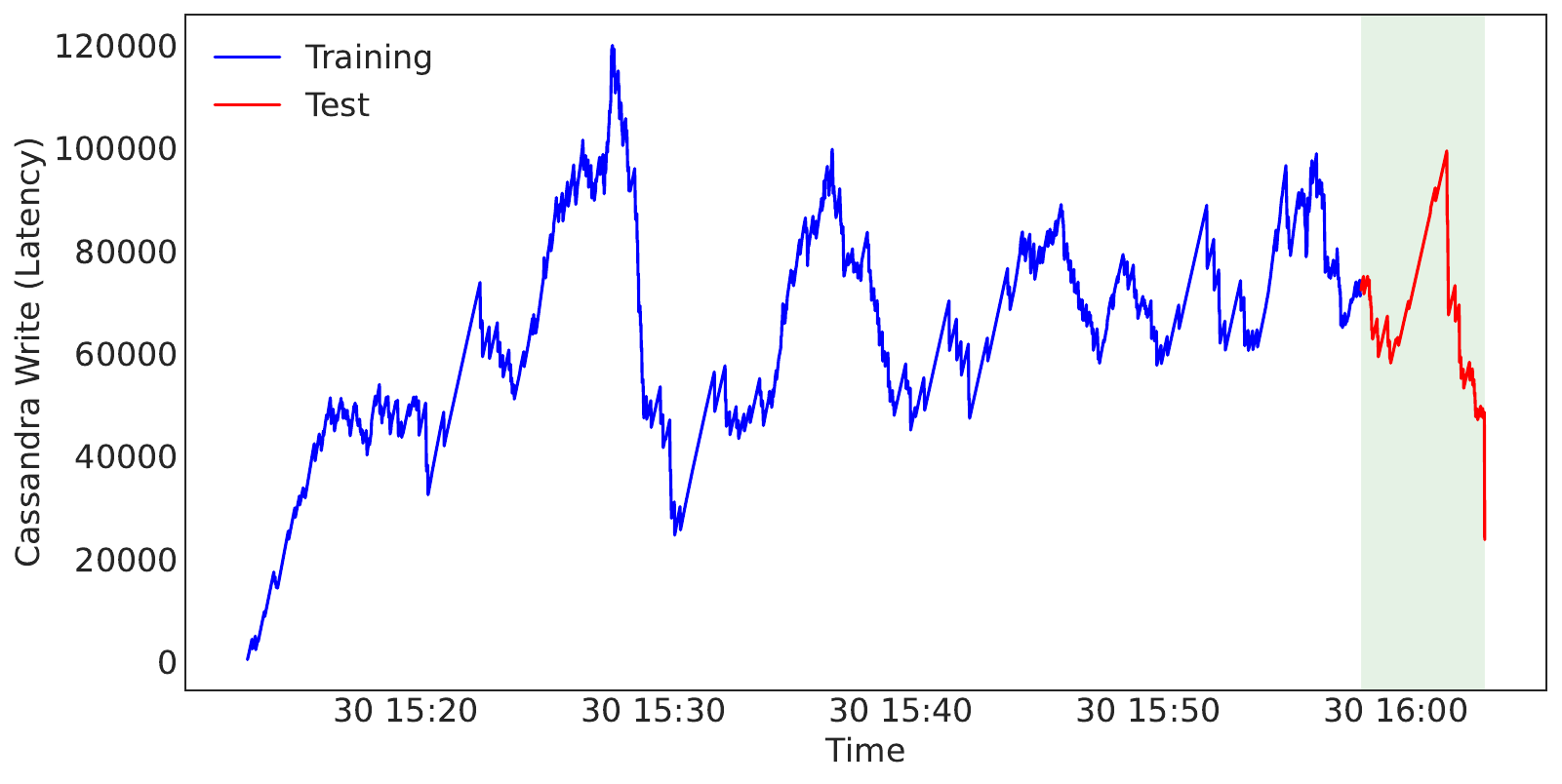}
    \caption{FIBRE-NG: Write Operation.}
    \label{fig:fibre_ng_write}
\end{figure}

\begin{figure}[htbp]
    \centering
    \includegraphics[width=0.9\columnwidth]{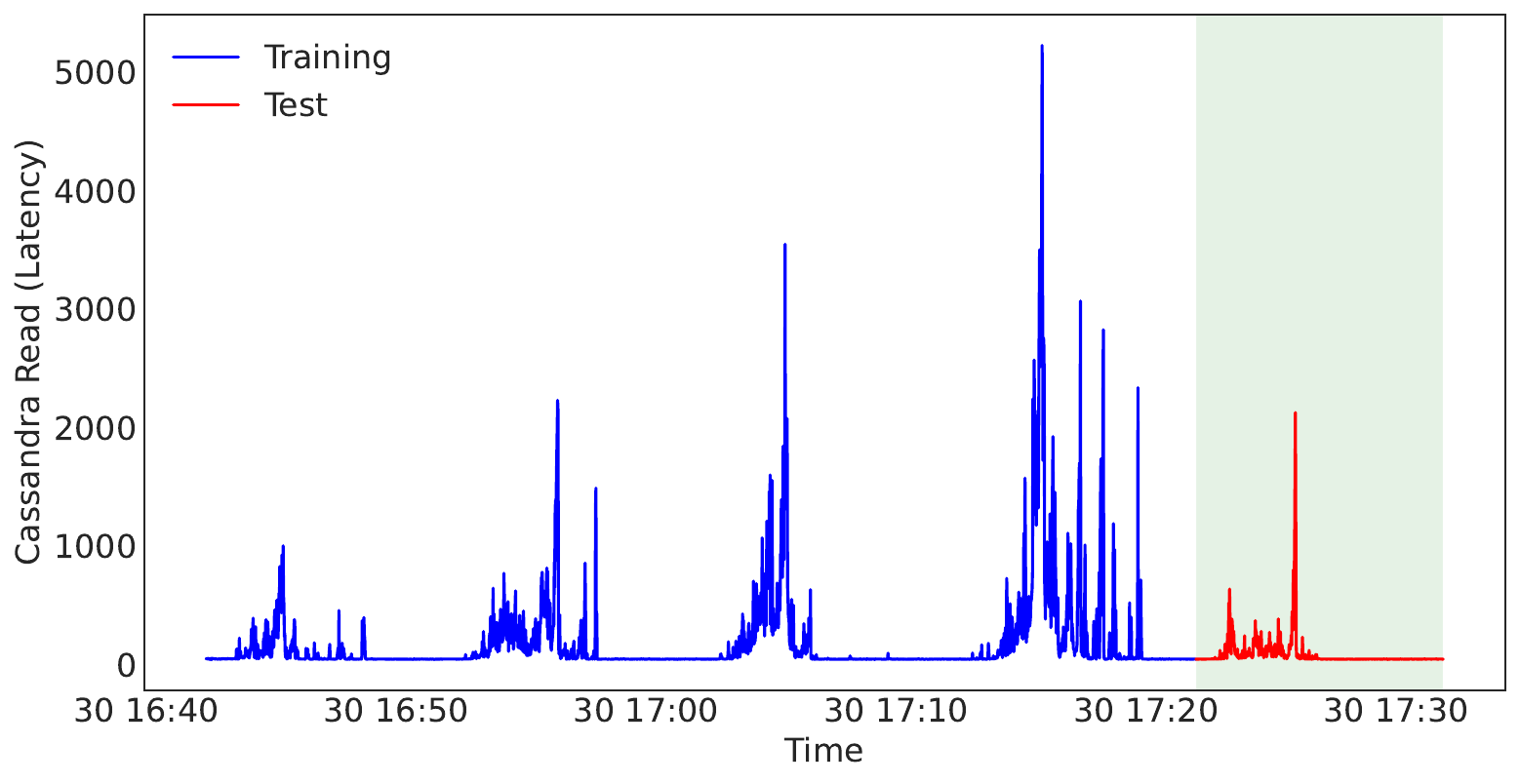}
    \caption{FIBRE-NG: Read Operation.}
    \label{fig:fibre_ng_read}
\end{figure}

\begin{figure}[htbp]
    \centering
    \includegraphics[width=0.9\columnwidth]{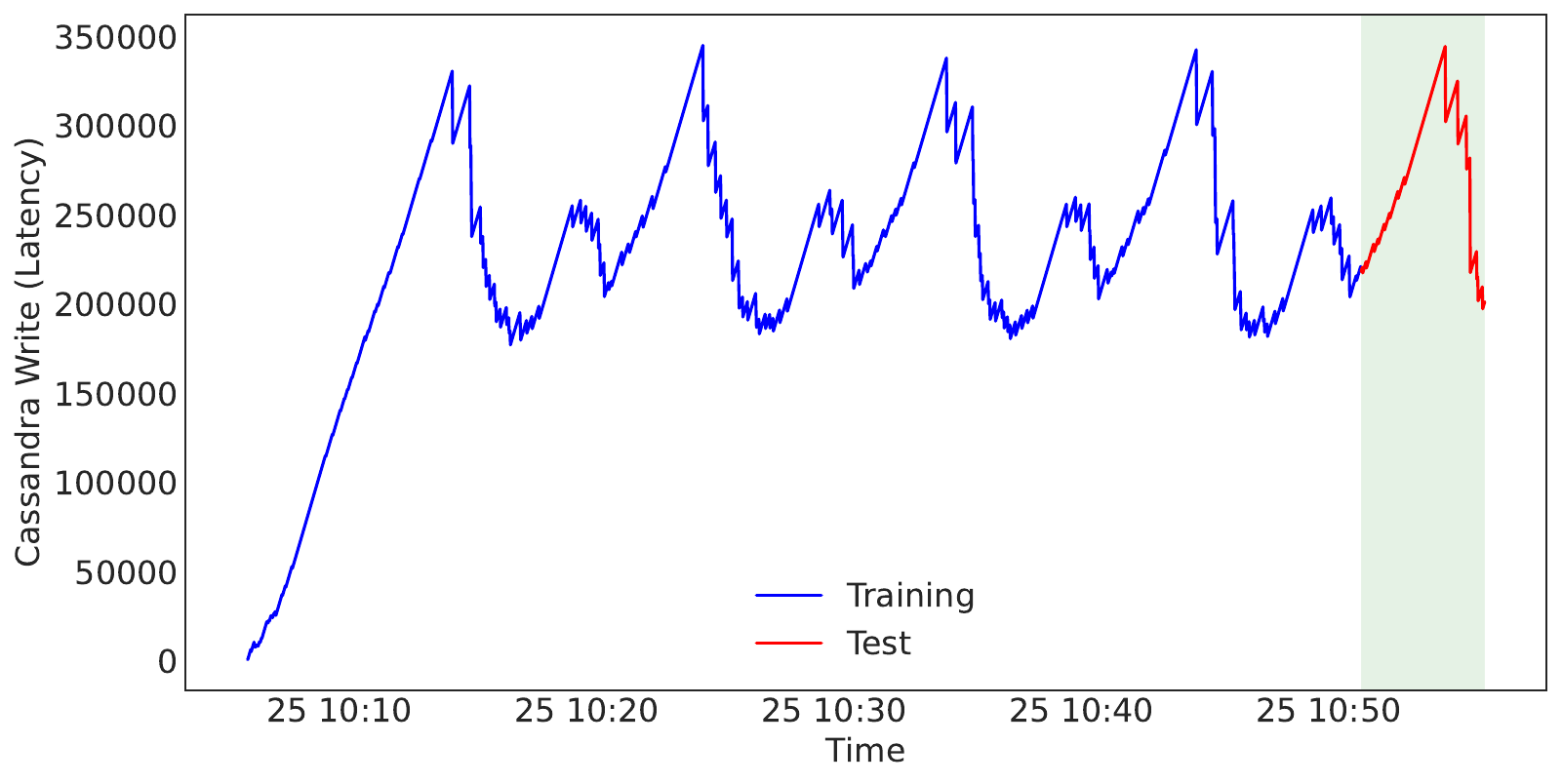}
    \caption{Fabric: Write Operation.}
    \label{fig:fabric_write}
\end{figure}

\begin{figure}[htbp]
    \centering
    \includegraphics[width=0.9\columnwidth]{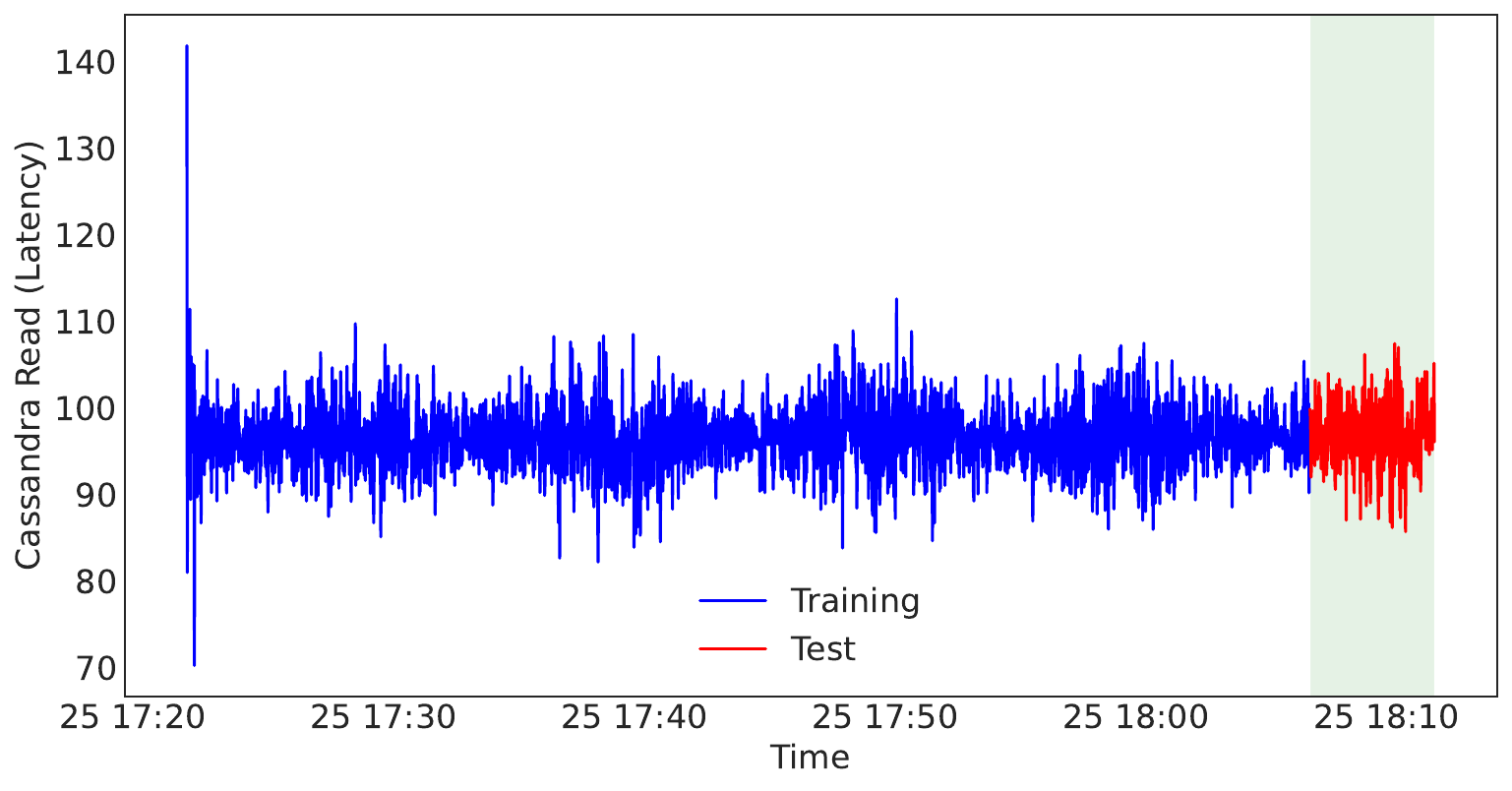}
    \caption{Fabric: Read Operation.}
    \label{fig:fabric_read}
\end{figure}

This method involves forecasting the next data point based on historical observations up to the current time and making continuous predictions as new data become available. The set of data generated from the \textbf{W}rite and \textbf{R}ead operations, as well as the training and testing splits, is shown in \textcolor{black}{Figs.~\ref{fig:fibre_ng_write}, ~\ref{fig:fibre_ng_read}, ~\ref{fig:fabric_write}, and~\ref{fig:fabric_read}}, where the operation latency is $ms$ on the y-axis, and the timestamp of the experiment is on the x-axis.

\textcolor{black}{To adapt our structured numerical dataset for \ac{DNNs}, specifically for \ac{CNNs}, we applied a sliding window transformation using the Sliding Window method from the \texttt{tsai} library. Given a dataset with $n$ variables (merged monitored metrics), where the last column represents the target variable (Cassandra latency), we constructed overlapping windows of length $w = 50$ with a stride of $s = 1$. Formally, let $\mathbf{X} \in \mathbb{R}^{m \times w \times (n-1)}$ be the input tensor and $\mathbf{y} \in \mathbb{R}^{m \times 1}$ be the corresponding target values, where $m$ is the number of generated windows. Each window $\mathbf{X}_i$ is defined as:}

\begin{equation}
\mathbf{X}_i = [\mathbf{x}_{i}, \mathbf{x}_{i+1}, \dots, \mathbf{x}_{i+w-1}], \quad \mathbf{y}_i = x_{i+w}^{\text{target}}
\end{equation}

\textcolor{black}{where $\mathbf{x}_j \in \mathbb{R}^{n-1}$ represents the feature values at time step $j$. This transformation allows CNNs to extract spatial and temporal patterns effectively, treating each window as a structured input similar to an image. The sliding mechanism ensures that local dependencies are captured while enabling the model to generalize across different time segments. To evaluate model performance, we employed a time-based split using the Time Splitter function, reserving a predefined number of samples for testing.}

\section{Experiments and Model Computation}\label{sec:experiments_and_model_computation}

Our experiments sought to validate the behavior of \ac{DNNs} in estimating \ac{SLA} compliance in a nationwide network slice deployed through the \ac{SFI2} Orchestrator. Initially, we deployed the application to the testbeds and started the workload tests to generate datasets for different metrics. \textcolor{black}{For each testbed, we seek to identify the Cassandra application's behavior and \ac{SLA} and understand whether \ac{DNNs} or basic \ac{ML} algorithms can generalize predictions across geographically distributed testbeds that experience produc\-tion-ready network conditions}.

\subsection{Testbeds}\label{subsec:testbeds}

\begin{figure*}[htbp]
  \centering
  \includegraphics[width=1.4\columnwidth]{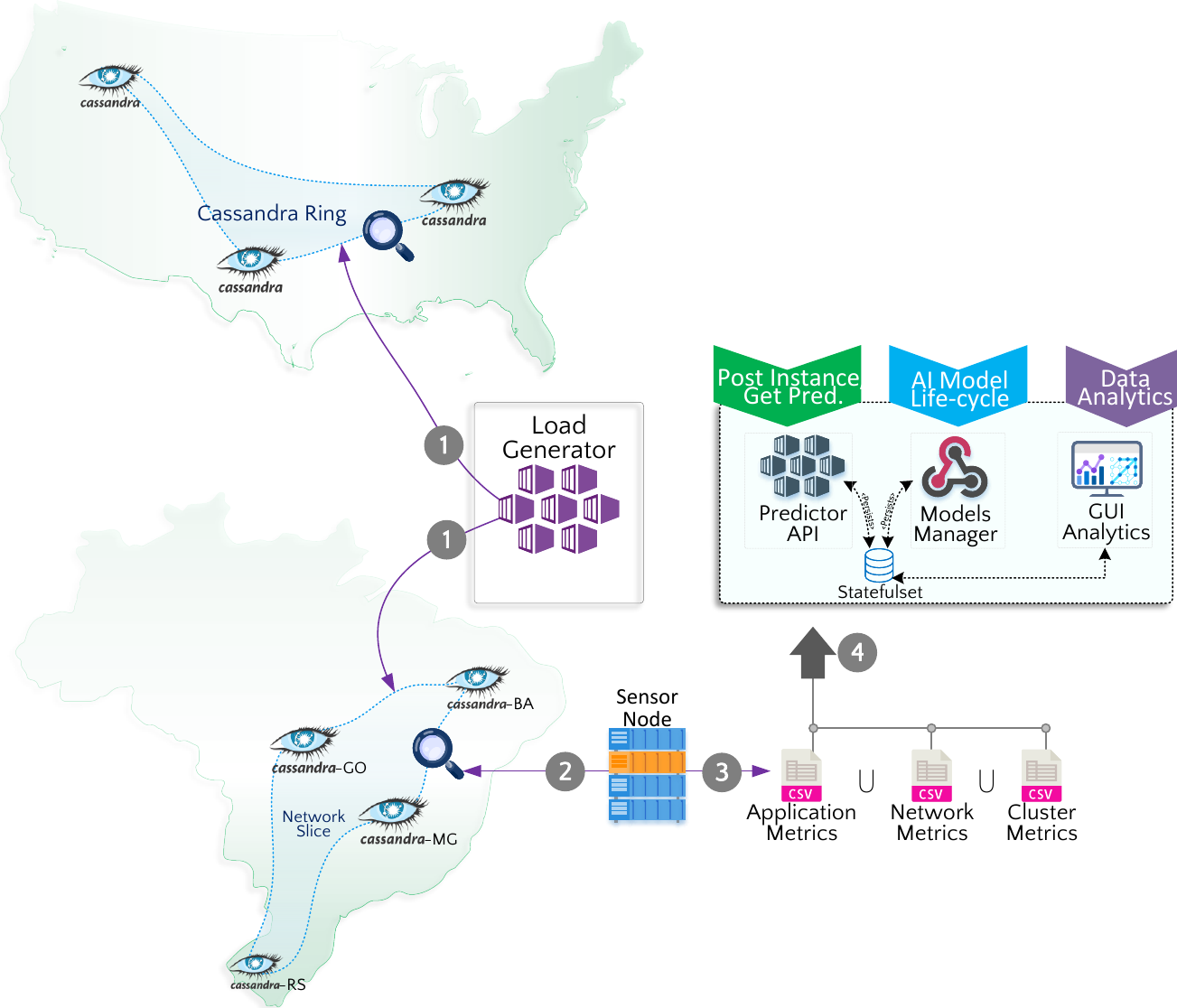}
  \caption{Deployment of Cassandra Application on nationwide testbeds.}
  \label{fig:experimental_flow_example}
\end{figure*}

Fig.~\ref{fig:experimental_flow_example} shows the experimental setup used for our evaluation. What stands out is the Cassandra ring deployed on different computing nodes is spread across each testbed \ac{FIBRE-NG}~\cite{salmito2014fibre} and Fabric~\cite{Baldin2019}. We collected and processed the monitoring metrics, which are the features $X = X_{application} \cup X_{cluster} \cup X_{network}$, feeding the Predictor \ac{API} that applies training with the different regression models.

To collect the application metrics, we employed a sensor and load-generating node. The load-generating pipeline shown in Fig.~\ref{fig:experimental_flow_example} executed read and write requests against the Cassandra ring (step \circlednumber{1}), whereas the sensor node generated flat requests simultaneously without altering the workload pattern (step \circlednumber{2}). Thus, we successfully gathered the necessary metrics of the application from the server side, which were subsequently compiled \textcolor{black}{(step \circlednumber{3}) into a dataset}. \textcolor{black}{Finally, the dataset is uploaded to the \ac{SLA} compliance training and validation framework (step \circlednumber{4})}.

We deployed our network slice with the Cassandra application on two nationwide testbeds, \ac{FIBRE-NG} and Fabric, considering the computational nodes in different geographic and intercontinental locations, such as the \ac{CERN} node, presented in Table~ \ref{tab:deployment_over_testbeds}, which shows the rest of each testbed where we deployed the Cassandra application. This experimental setup aims to validate how \ac{DNNs} deal with traces generated at nodes that transmit data in a production-ready network.

\begin{table}[htbp]
\centering
\caption{Testbeds Compute nodes hosting Cassandra services (containers) in a distributed manner.}
\label{tab:deployment_over_testbeds}
\resizebox{\columnwidth}{!}{%
\begin{tabular}{cccc}
\hline
\textbf{Testbed}                                       & \textbf{Pod Name} & \textbf{Node Name}                                & \textbf{Location}   \\ \hline
\multicolumn{1}{l}{\multirow{4}{*}{\textbf{FIBRE-NG}}} & cassandra-0       & WHX-SC                                            & Santa Catarina      \\
\multicolumn{1}{l}{}                                   & cassandra-1       & WHX-RS                                            & Rio Grande do Sul   \\
\multicolumn{1}{l}{}                                   & cassandra-2       & WHX-PB                                            & Paraíba             \\
\multicolumn{1}{l}{}                                   & loadgen           & WHX-RN                                            & Rio Grande do Norte \\ \hline
\multirow{4}{*}{\textbf{Fabric}}                       & cassandra-0       & \ac{GPN}                       & Kansas City, MO     \\
                                                       & cassandra-1       & \ac{CERN} & France               \\
                                                       & cassandra-2       & The University of Utah                            & Salt Lake City, UT  \\
                                                       & loadgen           & Rutgers University                                & Jersey Cirty, NJ    \\ \hline
\end{tabular}%
}
\end{table}
 

\subsection{Model Training}\label{subsec:model_training}

In the experimental training and validation phase of the \textcolor{black}{basic \ac{ML} algorithms and} \ac{DNNs}, we used RTX 4060ti 16 Gb GPU hardware with an \ac{CPU} Intel(R) Core(TM) i5-4430 CPU @ 3.00GHz with 32 GB of \ac{RAM}. Furthermore, we used the PyTorch framework together with \textit{FastAi}~\cite{Howard2020} and \textit{tsai}~\cite{tsai} tools to build the \ac{DNNs} training setup. 
All models were trained \textcolor{black}{and validated with no seed locks} and with tuned hyperparameters, and we collected ten (10) samples of training time and performance metrics. The choice of \ac{DNNs} and \textcolor{black}{basic \ac{ML} models} aims to represent different model architectures for generalization and in-depth analysis.

\textcolor{black}{Our training and test pipeline involves data ingestion, timestamping, and indexing, followed by splitting into training and testing subsets. Data transformation includes resampling, interpolation, and normalization, with feature selection and separate scaling of the target variable. The Sliding Window method generates input-output pairs for time series forecasting, while TSDatasets and TSDataLoaders apply further transformations. Hyperopt optimizes hyperparameters like batch size and learning rate. The model is trained using Learner with early stopping to prevent overfitting, and performance metrics are recorded for evaluation, ensuring robust data preparation and optimization.}

\section{Evaluating Results}\label{sec:evaluation_results}

\textcolor{black}{To validate our contribution, we initially analyze the impact of different network factors on our case study network slicing application response time. We also discuss aspects of Model Tuning for our employed neural network. Later, we evaluate different \ac{ML} approaches on a production-ready network and assess the feasibility of our dataset generation method, which expresses real network conditions and enables fitting and training models to forecast \ac{SLA} violations on network slicing architectures.}

\subsection{Impact Analysis in a Large-Scale Network}\label{subsec:influence_analyses}
\textcolor{black}{To assess the impact of realistic and production-ready network metrics on our experimental deployment, we used a chaos engineering tool to simulate latency with jitter and packet loss~\cite{chaosmesh}. We systematically applied jitter ranging from 1 ms to 10 ms, following a uniform distribution, and introduced a packet loss between 1\% and 10\% in our experimental sliced application. For token management, we implemented two distinct slices containing our distributed database application by varying the number of tokens from 32 to 256.}

\textcolor{black}{We employed three-way \ac{ANOVA} to evaluate the effects of three independent factors and their interactions on a dependent variable. The factors are: (1) Fixed Latency with Jitter induction with levels 1ms and 10ms, (2) Network Packet Loss with levels 1\% and 10\%, and (3) Cassandra Tokens with levels 32 and 256. The response variable analyzed is the Cassandra Write and Read Latency. This method tests the main effects of each factor, the two-way interactions between factors, and the three-way interaction, using the F-statistic and corresponding p-value. A p-value of $p < 0.05$ indicates that the effect or interaction is statistically significant.}

\begin{table}[h!]
\centering
\caption{\textcolor{black}{Experimental Factors influence on Write Operations.}}
\resizebox{\columnwidth}{!}{%
\begin{tabular}{lcc}
\hline
\textbf{Source of Variation} & \textbf{F Value} & \textbf{P Value} \\ \hline
\textit{Network Delay} & 7.58245 & 0.00612 \\
\textit{Network Loss} & 580.45267 & $< 0.0001$ \\
\textit{Cassandra Tokens} & 0.25329 & 0.615 \\
\textit{Network Delay $\times$ Network Loss} & 4.32445 & 0.03811 \\
\textit{Network Delay $\times$ Cassandra Tokens} & 0.45758 & 0.49909 \\
\textit{Network Loss $\times$ Cassandra Tokens} & 1.24777 & 0.26455 \\
\textit{Network Delay $\times$ Network Loss $\times$ Cassandra Tokens} & 1.77435 & 0.18349 \\ \hline
\end{tabular}%
}
\label{tab:write_anova_clean}
\end{table}

\textcolor{black}{The three-way \ac{ANOVA} results as Table~\ref{tab:write_anova_clean} show that for Write operations, Network Delay (Jitter) ($F(1,472) = 7.58$, $p = 0.00612$) and Network Packet Loss ($F(1,472) = 580.45$, $p < 0.0001$) have significant effects on latency, while Cassandra Tokens ($F(1,472) = 0.25$, $p = 0.615$) do not. The interaction between Network Delay and Network Packet Loss is also significant ($F(1,472) = 4.32$, $p = 0.03811$), suggesting non-additive effects. However, interactions involving Cassandra Tokens and the three-way interaction are not significant ($p > 0.05$). These results emphasize that network-related factors are the primary contributors to latency during Write operations.}

\begin{figure}
  \centering
  \includegraphics[width=0.8\columnwidth]{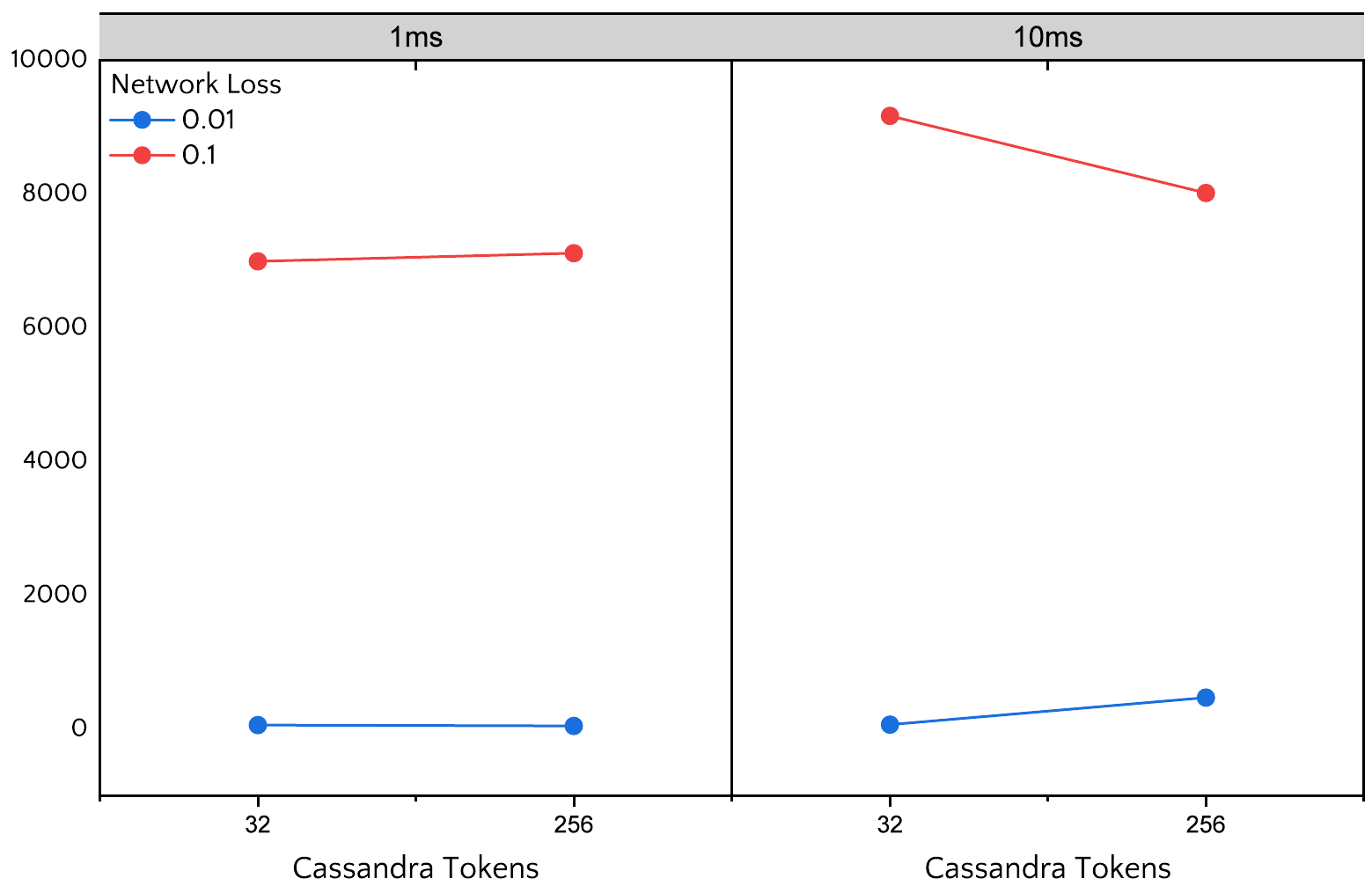}
  \caption{\textcolor{black}{Impact of Network Latency, Cassandra Tokens and Packet Loss on Cassandra Write Operations.}}
  \label{fig:anova_write_figure}
\end{figure}

\textcolor{black}{The results in Figura~\ref{fig:anova_write_figure} that high network latency (10ms) combined with high packet loss (0.1) significantly degrades performance, as seen in the red line dropping from 9000 to 8000. In contrast, low latency (1ms) with high packet loss starts at around 7000, with a slight increase. Configurations with low packet loss (0.01) maintain stable performance despite increased latency. ANOVA confirms these findings, with network delay and loss showing high statistical significance (P < 0.0001). These results highlight the importance of optimizing network conditions for enhancing the efficiency and reliability of distributed database systems like Cassandra.}

\textcolor{black}{The three-way \ac{ANOVA} results for Read operations as Table~\ref{tab:read_anova_clean} indicate that both Network Delay ($F(1,473) = 4.19$, $p = 0.04124$) and Network Packet Loss ($F(1,473) = 172.78$, $p < 0.0001$) significantly affect latency. Additionally, Cassandra Tokens shows a marginally significant effect ($F(1,473) = 3.81$, $p = 0.0515$), suggesting a potential influence on latency that may be worth further exploration. The interaction between Network Delay and Network Packet Loss is also nearly significant ($F(1,473) = 3.58$, $p = 0.05896$), while the interaction between Network Delay and Cassandra Tokens is not significant ($F(1,473) = 0.48$, $p = 0.491$). Furthermore, the interaction between Network Loss and Cassandra Tokens is significant ($F(1,473) = 5.79$, $p = 0.01648$), suggesting that these two factors jointly influence latency. The three-way interaction between Network Delay, Network Loss, and Cassandra Tokens is not significant ($F(1,473) = 2.69$, $p = 0.10178$). These findings highlight the dominant influence of network-related factors on latency during Read operations, with Cassandra Tokens potentially having a marginal effect and interactions between network factors playing a role.}

\begin{table}[h!]
\centering
\caption{\textcolor{black}{Experimental Factors influence on Read Operations.}}
\resizebox{\columnwidth}{!}{%
\begin{tabular}{lccc}
\hline
\textbf{Source of Variation} & \textbf{F Value} & \textbf{p Value} \\ \hline
\textit{Network Delay} & 4.19 & 0.04124 \\
\textit{Network Packet Loss} & 172.78 & $< 0.0001$ \\
\textit{Cassandra Tokens} & 3.81 & 0.0515 \\
\textit{Network Delay $\times$ Network Packet Loss} & 3.58 & 0.05896 \\
\textit{Network Delay $\times$ Cassandra Tokens} & 0.48 & 0.491 \\
\textit{Network Loss $\times$ Cassandra Tokens} & 5.79 & 0.01648 \\
\textit{Network Delay $\times$ Network Loss $\times$ Cassandra Tokens} & 2.69 & 0.10178 \\ \hline
\end{tabular}%
}
\label{tab:read_anova_clean}
\end{table}

\begin{figure}[H]
 \centering
  \includegraphics[width=0.8\columnwidth]{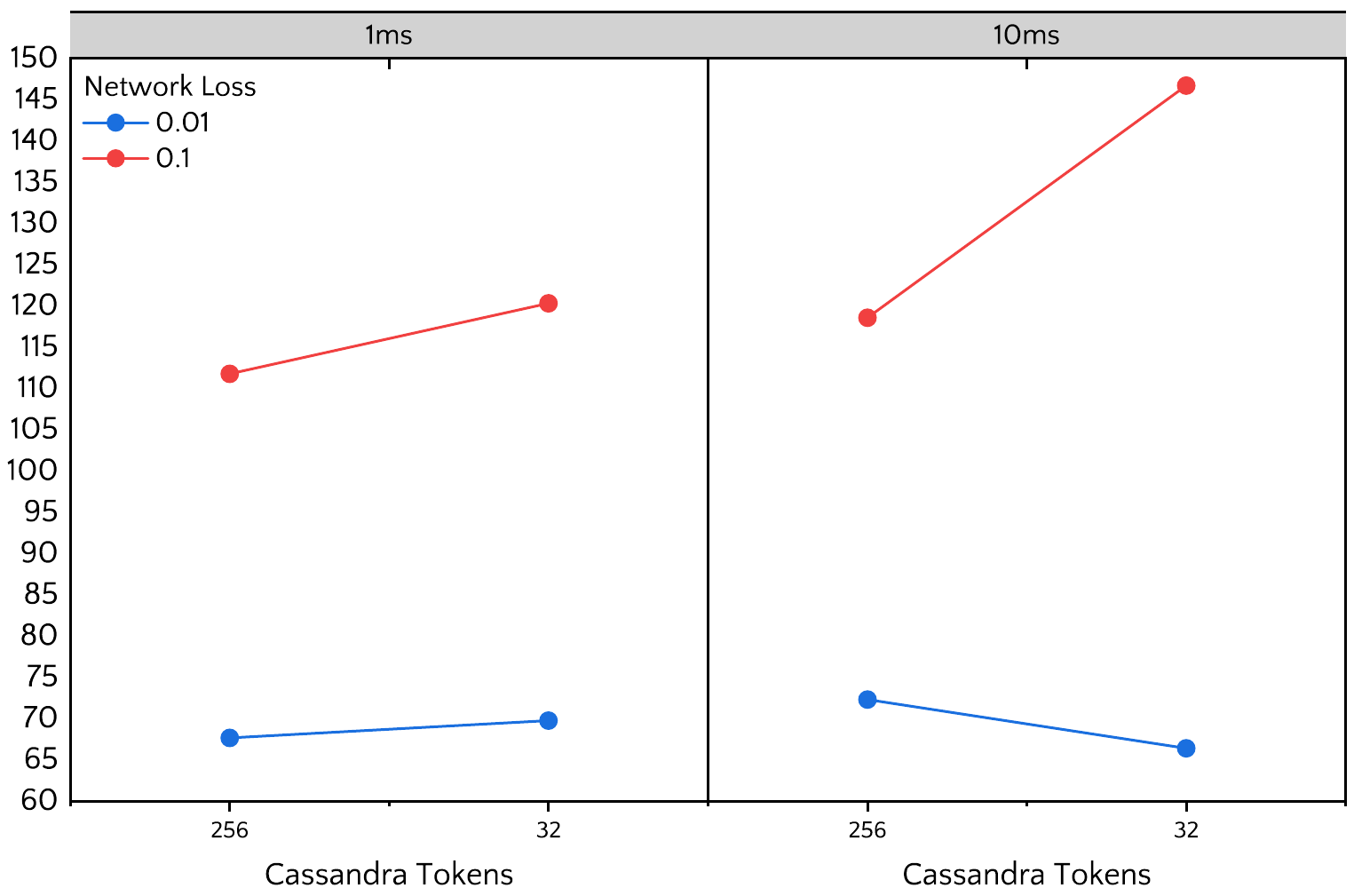}
  \caption{\textcolor{black}{Impact of Network Latency, Cassandra Tokens and Packet Loss on Cassandra Read Operations.}}
  \label{fig:anova_read_figure}
\end{figure}

\textcolor{black}{In Figure~\ref{fig:anova_read_figure} suggests that high network latency (10ms) combined with high packet loss (0.1) significantly degrades performance, with a notable decline observed in the graph. Conversely, lower latency (1ms) with high packet loss exhibits a slight performance increase. Moreover, configurations with low packet loss (0.01) demonstrate minimal impact from increased latency, indicating stability in performance. These trends are corroborated by \ac{ANOVA} results, highlighting the statistical significance of network delay (P = 0.00612) and network loss (P < 0.0001), as well as their interaction (P = 0.03811). These findings underscore the importance of optimizing both latency and packet loss to enhance the efficiency and reliability of distributed database systems like Cassandra.}

\subsection{Model Tunning}

In this section, we present the results of the hyperparameter optimization of \ac{DNNs} and the performance in estimating \ac{SLA} for each model for the different operations (W and R) and testbeds (\ac{FIBRE-NG} and Fabric). Therefore, we conducted hyperparameter optimization for the four constructed datasets \texttt{`fibre-read.csv}', \texttt{`fibre-write.csv'}, \texttt{`fabric-read.csv'}, and \texttt{`fabric-write.csv'}.

We summarize the hyperparameters found by Optuna for the \ac{FIBRE-NG} testbed in Table~\ref{tab:write_fibre-ng_hyperparameters_optimized}, and \ref{tab:read_fibre-ng_hyperparameters_optimized}. Furthermore, we tuned the hyperparameters of the same models for the dataset generated from the Fabric testbed, as shown in Table~\ref{tab:write_fabric_hyperparameters_optimized},~\ref{tab:read_fabric_hyperparameters_optimized}. With these adjusted values, we proceeded with an empirical evaluation of the performance of these models for predicting the \ac{SLA} in each testbed.

\begin{table}[H]
\caption{\textcolor{black}{Write Dataset: Hyperparameter tuning values for the FIBRE-NG Testbed.}}
\label{tab:write_fibre-ng_hyperparameters_optimized}
\renewcommand{\arraystretch}{1.6} 
\setlength{\tabcolsep}{2pt} 
\centering
\resizebox{\columnwidth}{!}{%
\begin{tabular}{lcccccccc}
\hline
\textbf{Model}             & \textbf{Batch Size} & \textbf{Learning Rate} & \textbf{Epochs} & \textbf{Patience} & \textbf{Optimizer} & \textbf{Layers} & \textbf{Hidden Size} & \textbf{Bidirectional} \\ \hline
\textbf{FCN}               & 8   & 0.1   & 20  & 10  & Adam & 2  & 100  & \textit{False}  \\
\textbf{FCNPlus}           & 16  & 0.1   & 20  & 50  & Adam & 4  & 200  & \textit{False}  \\
\textbf{InceptionTime}     & 16  & 0.1   & 100 & 50  & Adam & 4  & 100  & \textit{False}  \\
\textbf{InceptionTimePlus} & 16  & 0.1   & 100 & 50  & Adam & 1  & 50   & \textit{True}   \\
\textbf{OmniScaleCNN}      & 32  & 0.1   & 20  & 50  & Adam & 2  & 100  & \textit{True}   \\
\textbf{ResCNN}            & 32  & 0.1   & 20  & 10  & Adam & 5  & 100  & \textit{False}  \\
\textbf{ResNet}            & 32  & 0.1   & 50  & 50  & Adam & 2  & 100  & \textit{False}  \\
\textbf{ResNetPlus}        & 32  & 0.1   & 20  & 5   & Adam & 5  & 50   & \textit{False}  \\
\textbf{TCN}               & 32  & 0.01  & 100 & 50  & Adam & 1  & 200  & \textit{False}  \\
\textbf{XCM}               & 16  & 0.01  & 50  & 50  & Adam & 1  & 200  & \textit{False}  \\
\textbf{XCMPlus}           & 8   & 0.1   & 100 & 50  & Adam & 3  & 100  & \textit{True}   \\ \hline
\end{tabular}%
}
\end{table}

\begin{table}[H]
\caption{\textcolor{black}{Read Dataset: Hyperparameter tuning values for the FIBRE-NG Testbed.}}
\label{tab:read_fibre-ng_hyperparameters_optimized}
\renewcommand{\arraystretch}{1.6} 
\setlength{\tabcolsep}{2pt} 
\centering
\resizebox{\columnwidth}{!}{%
\begin{tabular}{lcccccccc}
\hline
\textbf{Model}             & \textbf{Batch Size} & \textbf{Learning Rate} & \textbf{Epochs} & \textbf{Patience} & \textbf{Optimizer} & \textbf{Layers} & \textbf{Hidden Size} & \textbf{Bidirectional} \\ \hline
\textbf{FCN}               & 16  & 0.1   & 20  & 10  & Adam & 3  & 200  & \textit{True}   \\
\textbf{FCNPlus}           & 16  & 0.1   & 20  & 50  & Adam & 4  & 50   & \textit{False}  \\
\textbf{InceptionTime}     & 8   & 0.01  & 20  & 10  & Adam & 5  & 100  & \textit{True}   \\
\textbf{InceptionTimePlus} & 16  & 0.1   & 20  & 50  & Adam & 2  & 200  & \textit{False}  \\
\textbf{OmniScaleCNN}      & 8   & 0.1   & 50  & 50  & Adam & 2  & 50   & \textit{True}   \\
\textbf{ResCNN}            & 16  & 0.1   & 20  & 10  & Adam & 3  & 100  & \textit{False}  \\
\textbf{ResNet}            & 16  & 0.1   & 20  & 10  & Adam & 1  & 200  & \textit{True}   \\
\textbf{ResNetPlus}        & 8   & 0.1   & 20  & 50  & Adam & 2  & 50   & \textit{False}  \\
\textbf{TCN}               & 32  & 0.01  & 50  & 5   & Adam & 4  & 50   & \textit{False}  \\
\textbf{XCM}               & 8   & 0.01  & 20  & 10  & Adam & 3  & 200  & \textit{False}  \\
\textbf{XCMPlus}           & 16  & 0.01  & 50  & 50  & Adam & 3  & 50   & \textit{True}   \\ \hline
\end{tabular}%
}
\end{table}

\begin{table}[H]
\caption{\textcolor{black}{Write Dataset: Hyperparameter tuning values for the Fabric Testbed.}}
\label{tab:write_fabric_hyperparameters_optimized}
\renewcommand{\arraystretch}{1.6} 
\setlength{\tabcolsep}{2pt} 
\resizebox{\columnwidth}{!}{%
\begin{tabular}{lcccccccc}
\hline
\textbf{Model}             & \textbf{Batch Size} & \textbf{Learning Rate (LR)} & \textbf{Epochs} & \textbf{Patience} & \textbf{Optimizer} & \textbf{\# of Layers} & \textbf{Hidden Size} & \textbf{Bidirectional} \\ \hline
\textbf{FCN}               & 32                  & 0.01                        & 20              & 10                & Adam               & 3                     & 50                   & True                   \\
\textbf{FCNPlus}           & 32                  & 0.1                         & 100             & 50                & Adam               & 5                     & 100                  & True                   \\
\textbf{InceptionTime}     & 32                  & 0.1                         & 50              & 50                & Adam               & 2                     & 200                  & False                  \\
\textbf{InceptionTimePlus} & 16                  & 0.1                         & 100             & 50                & Adam               & 4                     & 50                   & True                   \\
\textbf{OmniScaleCNN}      & 16                  & 0.01                        & 50              & 10                & Adam               & 4                     & 100                  & True                   \\
\textbf{ResCNN}            & 32                  & 0.01                        & 50              & 50                & Adam               & 5                     & 100                  & False                  \\
\textbf{ResNet}            & 32                  & 0.01                        & 50              & 50                & Adam               & 4                     & 50                   & True                   \\
\textbf{ResNetPlus}        & 16                  & 0.1                         & 20              & 10                & Adam               & 3                     & 200                  & True                   \\
\textbf{TCN}               & 8                   & 0.001                       & 100             & 50                & Adam               & 2                     & 50                   & False                  \\
\textbf{XCM}               & 16                  & 0.1                         & 50              & 50                & Adam               & 3                     & 50                   & False                  \\
\textbf{XCMPlus}           & 16                  & 0.1                         & 100             & 50                & Adam               & 1                     & 50                   & True                   \\ \hline
\end{tabular}%
}
\end{table}

\begin{table}[H]
\caption{\textcolor{black}{Read Dataset: Hyperparameter tuning values for the Fabric Testbed.}}
\label{tab:read_fabric_hyperparameters_optimized}
\renewcommand{\arraystretch}{1.6} 
\setlength{\tabcolsep}{2pt} 
\resizebox{\columnwidth}{!}{%
\begin{tabular}{lcccccccc}
\hline
\textbf{Model}             & \textbf{Batch Size} & \textbf{LR}  & \textbf{Epochs} & \textbf{Patience} & \textbf{Optimizer} & \textbf{\# Layers} & \textbf{Hidden Size} & \textbf{Bidirectional} \\ \hline
\textbf{FCN}               & 32                  & 0.1          & 20              & 10                & Adam               & 3                   & 100                  & \textit{False}         \\
\textbf{FCNPlus}           & 8                   & 0.1          & 20              & 10                & Adam               & 3                   & 200                  & \textit{True}          \\
\textbf{InceptionTime}     & 32                  & 0.1          & 100             & 10                & Adam               & 5                   & 50                   & \textit{True}          \\
\textbf{InceptionTimePlus} & 16                  & 0.1          & 100             & 50                & Adam               & 2                   & 50                   & \textit{False}         \\
\textbf{OmniScaleCNN}      & 8                   & 0.001        & 20              & 50                & SGD                & 4                   & 100                  & \textit{False}         \\
\textbf{ResCNN}            & 32                  & 0.1          & 20              & 10                & SGD                & 2                   & 100                  & \textit{True}          \\
\textbf{ResNet}            & 32                  & 0.1          & 20              & 50                & Adam               & 4                   & 100                  & \textit{True}          \\
\textbf{ResNetPlus}        & 8                   & 0.1          & 20              & 5                 & Adam               & 4                   & 200                  & \textit{False}         \\
\textbf{TCN}               & 32                  & 0.01         & 20              & 10                & Adam               & 2                   & 100                  & \textit{True}          \\
\textbf{XCM}               & 8                   & 0.1          & 100             & 10                & SGD                & 3                   & 50                   & \textit{True}          \\
\textbf{XCMPlus}           & 8                   & 0.1          & 100             & 10                & SGD                & 5                   & 200                  & \textit{True}          \\ \hline
\end{tabular}%
}
\end{table}



\subsection{Basic Model Performance}\label{subsec:basic_model_performance}

Through our dataset construction framework on sliced testbeds, we aggregate the metrics $X = X_{application} \cup X_{cluster} \cup X_{network}$ $\rightarrow$ $Y$ and configure the datasets generated to submit it to the training and assessment process. Thus, we empirically use the division \textcolor{black}{$80\%$ for training and $20\%$} for testing and build the regression model aiming at the one-step-ahead prediction method that, given a statistic $i$ of $X_{i}$, the model can estimate the operation latency (W or R) in the next step.

\begin{figure}[H]
    \centering
    \subfloat[\centering \scriptsize{\textcolor{black}{FIBRE-NG: MAPE (Write).}}]{
        \includegraphics[width=0.8\columnwidth]{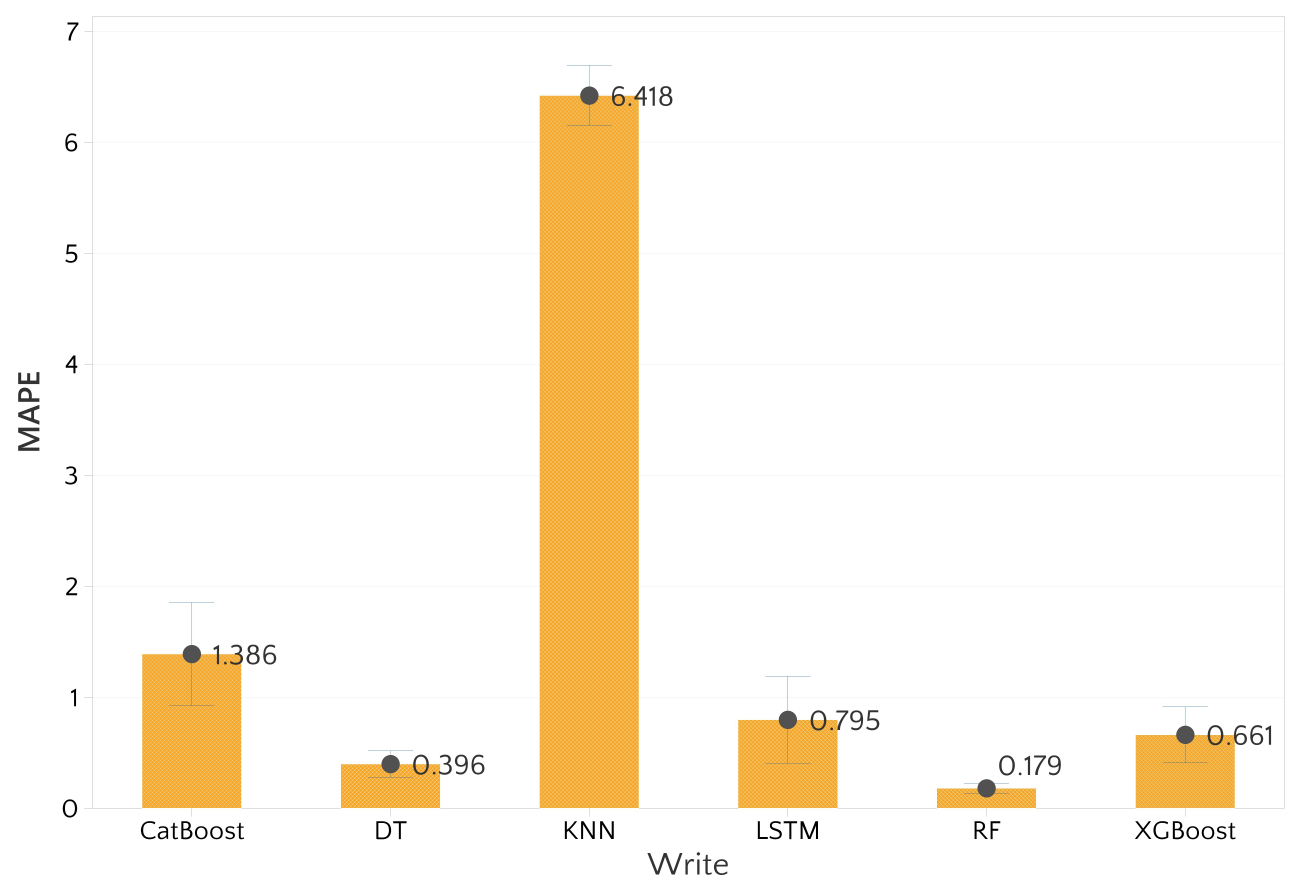}
    }
    \\
    \subfloat[\centering \scriptsize{\textcolor{black}{FIBRE-NG: MAPE (Read).}}]{
        \includegraphics[width=0.8\columnwidth]{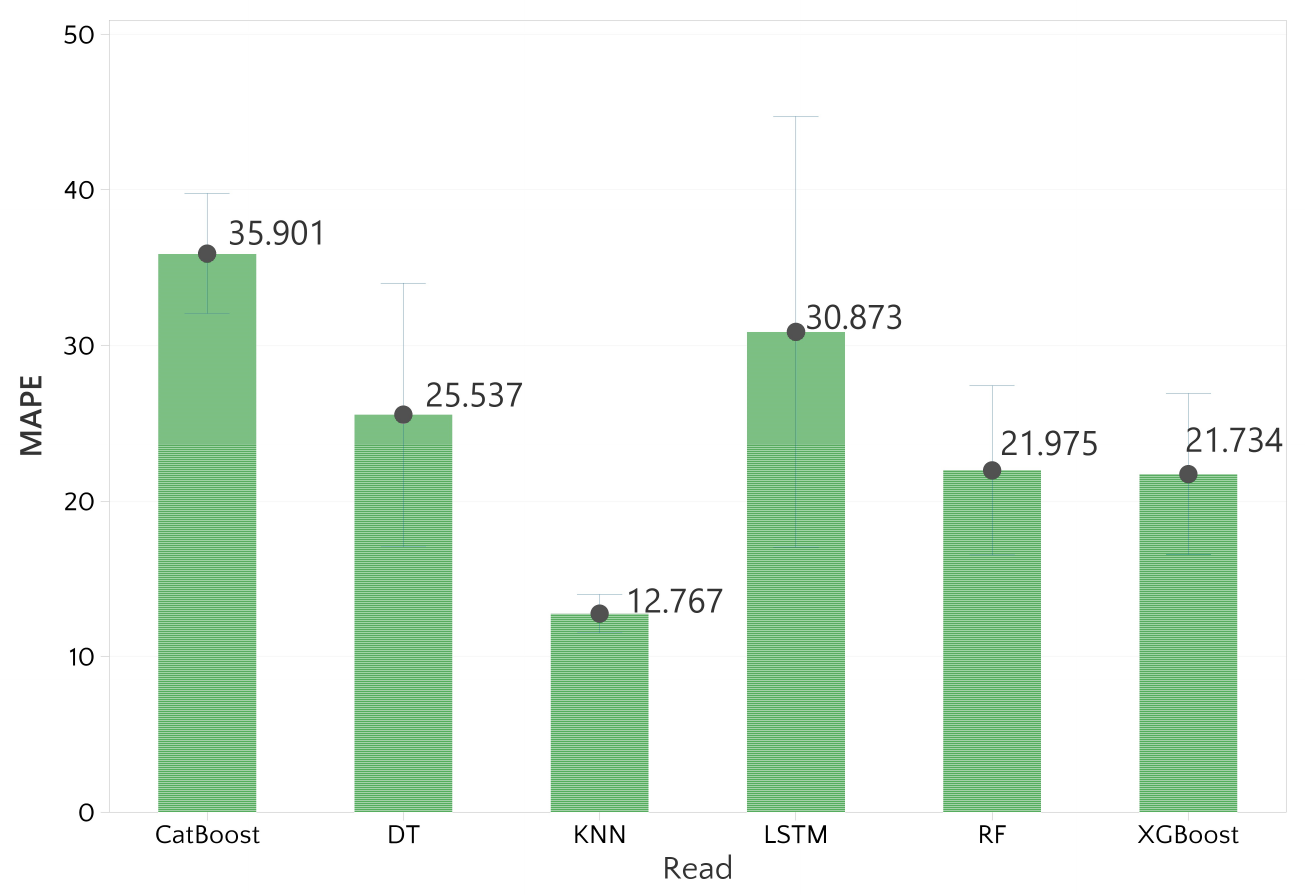}
    }
    \caption{\textcolor{black}{MAPE results for Write and Read operations on FIBRE-NG.}}
    \label{fig:FIBRE-NG_MAPE}
\end{figure}

\textcolor{black}{As Fig.~\ref{fig:FIBRE-NG_MAPE}-a, the Write operation on the FIBRE-NG testbed, and \ac{RF} presented the lowest mean \ac{MAPE} (0.17), indicating the best predictive accuracy, followed by \ac{DT} (0.39). The \ac{KNN} had the worst performance, with an average \ac{MAPE} of 6.41, which is significantly higher than that of the others. Models such as CatBoost, \ac{LSTM}, and XGBoost have intermediate performance but are still superior to \ac{KNN}. In addition, \ac{RF} demonstrated greater stability, with the lowest standard deviation (0.064).}

\textcolor{black}{In the Read operation, as shown in Fig.~\ref{fig:FIBRE-NG_MAPE}-b on the FIBRE-NG testbed, \ac{KNN} obtained the lowest mean \ac{MAPE} (12.77), indicating the best predictive accuracy. In contrast, CatBoost (35.90) and \ac{LSTM} (30.87) presented the highest errors with high variability, as evidenced by the high standard deviation of \ac{LSTM} (19.34). \ac{RF} and XGBoost had intermediate performances, with mean MAPEs of 21.98 and 21.73, respectively. Meanwhile, \ac{DT} showed a mean error of 25.54, with high dispersion.}

\begin{figure}[H]
    \centering
    \subfloat[\centering \scriptsize{\textcolor{black}{Fabric: MAPE (Write).}}]{
        \includegraphics[width=0.9\columnwidth]{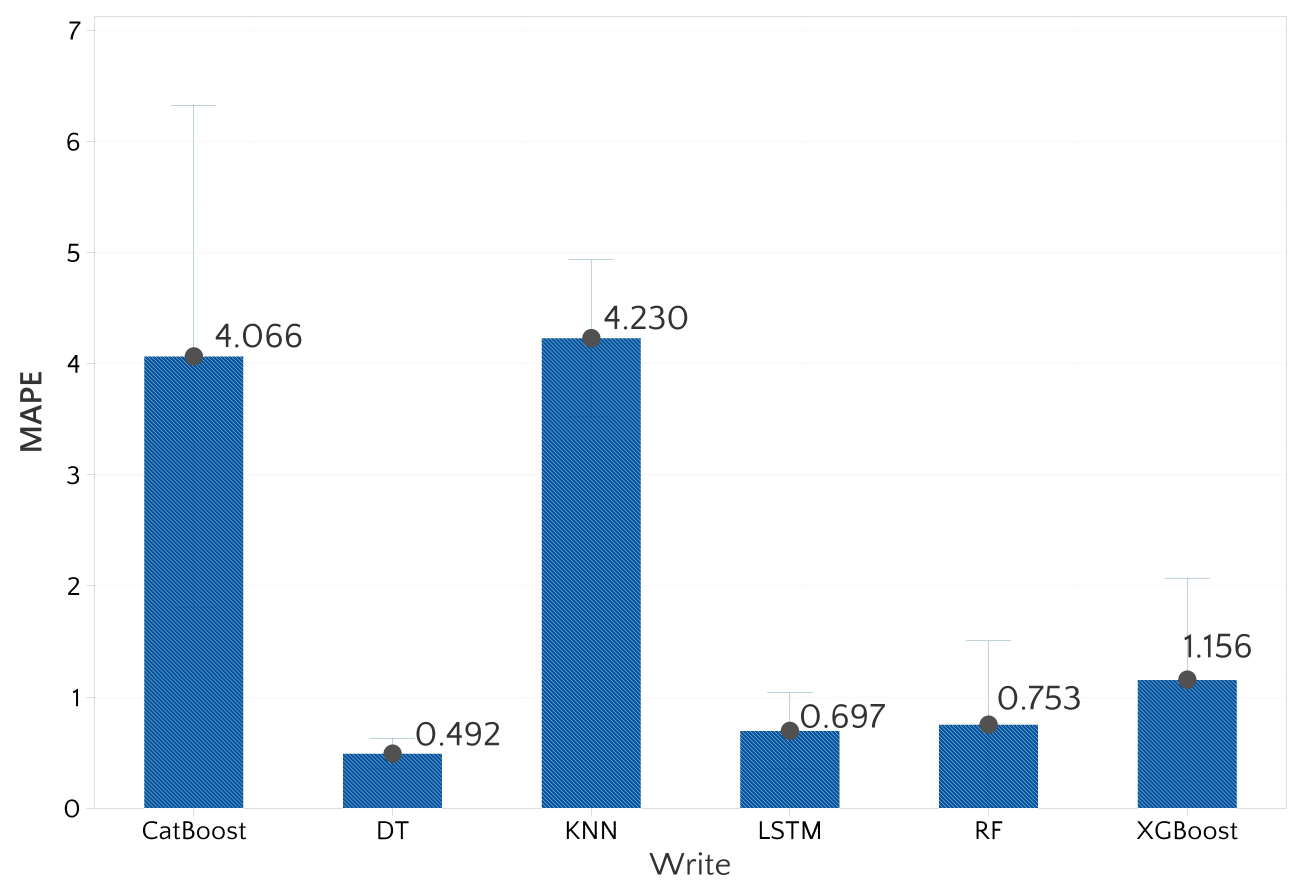}
    }
    \\
    \subfloat[\centering \scriptsize{\textcolor{black}{Fabric: MAPE (Read).}}]{
        \includegraphics[width=0.9\columnwidth]{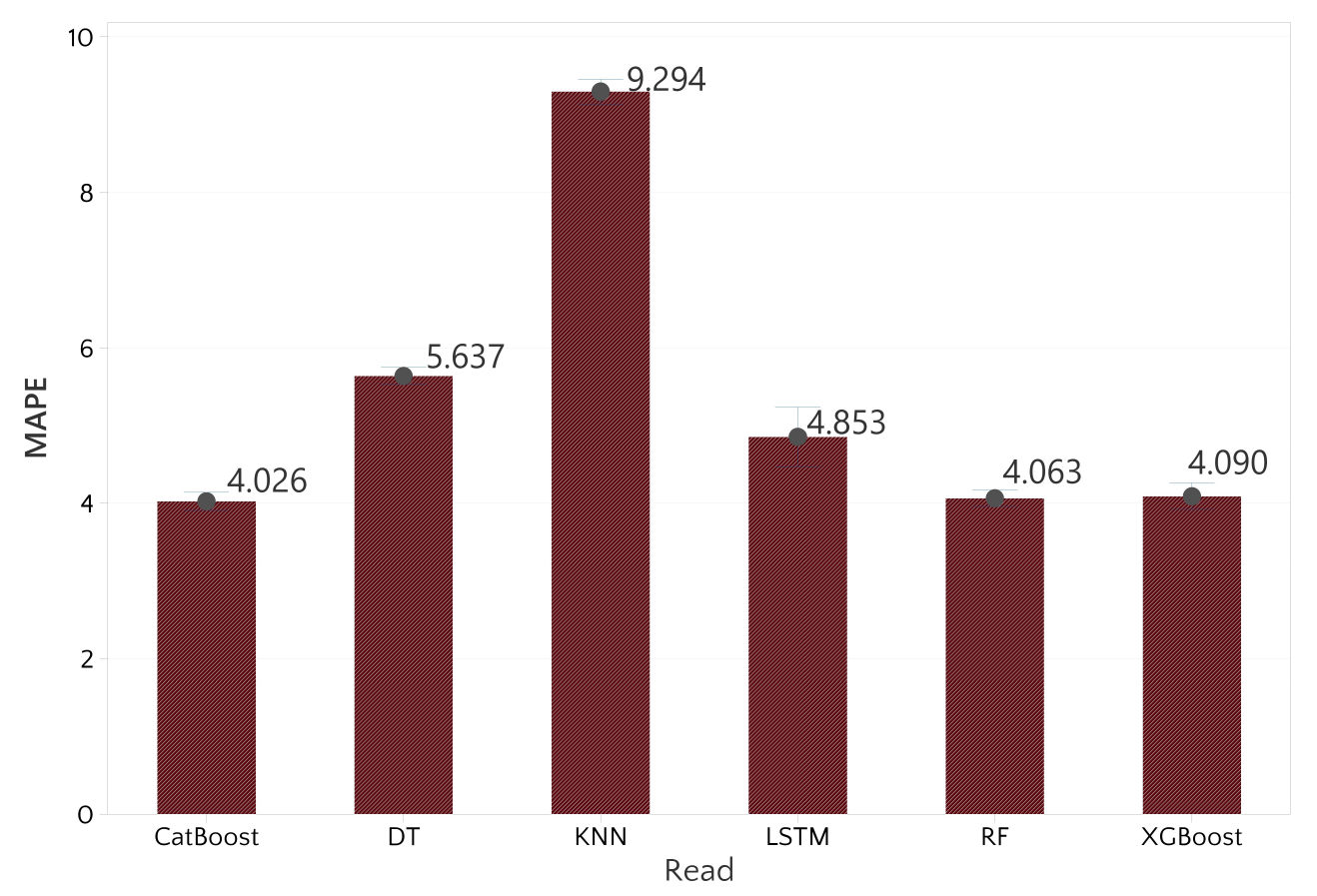}
    }
    \caption{\textcolor{black}{MAPE results for Write and Read operations on Fabric.}}
    \label{fig:Fabric_MAPE}
\end{figure}

\textcolor{black}{During the writing operation, as shown in Fig.~\ref{fig:Fabric_MAPE}-a, on the testbed Fabric, the \ac{DT} presented the lowest mean \ac{MAPE} (0.49), indicating the best predictive accuracy, followed by \ac{LSTM} (0.69) and \ac{RF} (0.75). XGBoost had a slightly higher error (1.16) but was still lower than CatBoost (4.07) and \ac{KNN} (4.23), which had the worst performance. In addition, \ac{DT} showed the lowest standard deviation (0.19), suggesting greater stability.}

\textcolor{black}{In the Read operation, as Fig.~\ref{fig:Fabric_MAPE}-b, on the testbed Fabric, \ac{RF} presented the lowest mean \ac{MAPE} (4.06), closely followed by XGBoost (4.09) and CatBoost (4.03), indicating very similar performances. \ac{DT} had a higher error (5.64), while \ac{KNN} obtained the worst result (9.29), with the highest standard deviation (0.22), demonstrating low precision. \ac{LSTM} showed an intermediate \ac{MAPE} (4.85), but with greater variability (0.53). Thus, RF, XGBoost, and CatBoost stand out as the most effective approaches for predicting the latency in the Read operation of the distributed Cassandra on the testbed Fabric.}

\textcolor{black}{These experiments with basic \ac{ML} algorithms for forecasting highlight the contribution of our study, which discusses the suitability of laboratory-trained algorithms in real-world scenarios and production networks with geographically distributed nodes. In the four analyzed scenarios, \ac{RF} and \ac{DT} demonstrated the lowest mean \ac{MAPE}, indicating higher predictive accuracy. In the Write operation on the Fabric testbed, \ac{DT} had the best performance (0.49), while in the Read operation on the same testbed, \ac{RF} stood out (4.06). On FIBRE-NG, \ac{RF} and \ac{KNN} were more accurate in the Read operation (21.98 and 12.77, respectively), whereas \ac{DT} had the lowest error in the Write operation (0.40). In contrast, \ac{KNN} and CatBoost exhibited the worst performances in various scenarios. These results suggest that tree-based models, especially \ac{RF} and \ac{DT}, are more effective for forecasting in sliced testbeds.}

\subsection{DNN-based Model Performance}\label{subsec:dnn_model_performance}

\begin{figure}[H]
  \centering
  \includegraphics[width=1\columnwidth]{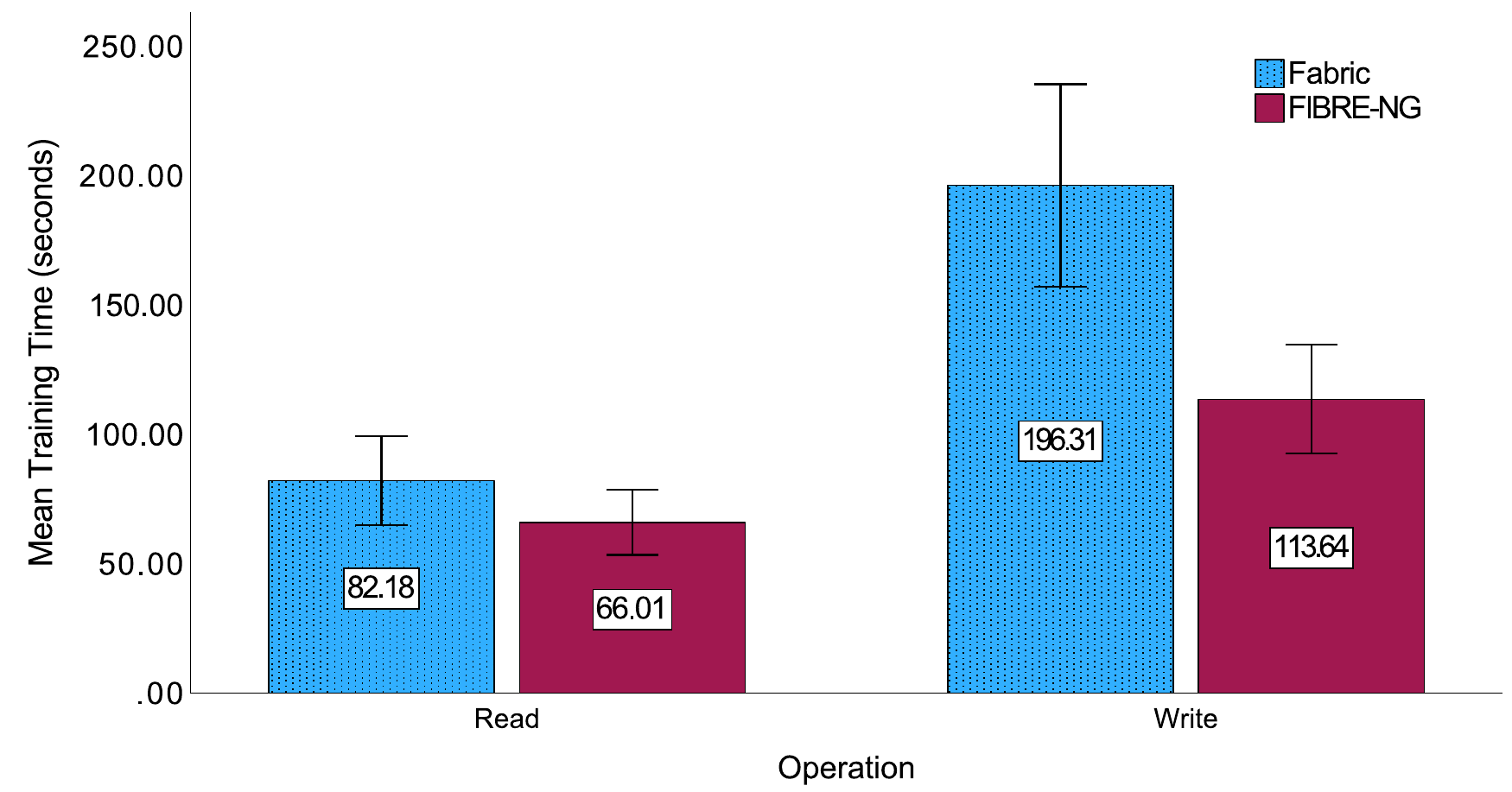}
  \caption{\textcolor{black}{Training Time for different Operations (W and R) on Testbeds.}}
  \label{fig:training_time_demands}
\end{figure}

We computed the \ac{DNNs} training time demands for \textbf{W}rite and \textbf{R}ead operations on the \ac{FIBRE-NG} and Fabric testbeds and compared the results in Fig.~\ref{fig:training_time_demands}. Therefore, it is possible to infer that with a confidence level of $95\%$, the read datasets generated in different testbeds require the same training time. In contrast, we noticed a difference in the training time between the Read and Write operations owing to the size of the generated dataset and the variations in network conditions experienced by each test workload.

We deepened our analysis by observing the generalization capacity of the \ac{DNNs} used to predict the Write and Read operations latency in Cassandra deployed on the \ac{FIBRE-NG} and Fabric Testbeds. We present in Fig.~\ref{fig:training_time_demands} the training and validation behaviors of only the best \ac{DNNs} for the two operations (W and R) and testbeds considering the \ac{MAE} metric and the measured values in Table~\ref{tab:fabric_mae} and Table~\ref{tab:fibre_mae}.

\begin{table}[H]
\centering
\caption{Summarizing the \ac{DNNs} performance for FABRIC using \ac{MAE}.}
\label{tab:fabric_mae}
\resizebox{\columnwidth}{!}{%
\begin{tabular}{cclll}
\hline
\textbf{Neural Network} & \textbf{Read Mean} & \textbf{Read StDev} & \textbf{Write Mean} & \textbf{Write StDev} \\ \hline
\textbf{FCN}           & 0.040              & 0.000               & 0.016               & 0.005               \\
\textbf{FCNPlus}      & 0.042              & 0.004               & 0.012               & 0.004               \\
\textbf{InceptionTime} & 0.041              & 0.003               & 0.020               & 0.008               \\
\textbf{InceptionTimePlus} & 0.040          & 0.000               & 0.030               & 0.017               \\
\textbf{OmniScaleCNN}  & 0.046              & 0.005               & 0.027               & 0.018               \\
\textbf{ResCNN}        & 0.043              & 0.005               & 0.021               & 0.007               \\
\textbf{ResNet}        & 0.040              & 0.000               & 0.059               & 0.033               \\
\textbf{ResNetPlus}    & 0.048              & 0.006               & 0.024               & 0.007               \\
\textbf{TCN}           & 0.040              & 0.000               & 0.012               & 0.004               \\
\textbf{XCM}           & 0.054              & 0.014               & 0.027               & 0.018               \\ \hline
\end{tabular}%
}
\end{table}

\begin{table}[H]
\centering
\caption{Summarizing the \ac{DNNs} performance for FIBRE-NG using \ac{MAE}.}
\label{tab:fibre_mae}
\resizebox{\columnwidth}{!}{%
\begin{tabular}{cclll}
\hline
\textbf{Neural Network} & \textbf{Read Mean} & \textbf{Read StDev} & \textbf{Write Mean} & \textbf{Write StDev} \\ \hline
\textbf{FCN}           & 0.017              & 0.008               & 0.021               & 0.007               \\
\textbf{FCNPlus}      & 0.013              & 0.005               & 0.014               & 0.005               \\
\textbf{InceptionTime} & 0.013              & 0.005               & 0.024               & 0.027               \\
\textbf{InceptionTimePlus} & 0.011          & 0.003               & 0.016               & 0.007               \\
\textbf{OmniScaleCNN}  & 0.020              & 0.009               & 0.035               & 0.025               \\
\textbf{ResCNN}        & 0.021              & 0.019               & 0.026               & 0.013               \\
\textbf{ResNet}        & 0.010              & 0.000               & 0.012               & 0.004               \\
\textbf{ResNetPlus}    & 0.014              & 0.005               & 0.017               & 0.005               \\
\textbf{TCN}           & 0.019              & 0.007               & 0.015               & 0.005               \\
\textbf{XCM}           & 0.025              & 0.007               & 0.035               & 0.012               \\ \hline
\end{tabular}%
}
\end{table}

We choose to employ the \ac{MAE} metric rather than \ac{RMSE} or \ac{MSE} for regression on the Read and Write latency owing to its robustness to outlier handling and ease of interpretation. Additionally, \ac{MAE} shares the same unit of measurement as the dependent variable, facilitating a more intuitive comprehension of its meaning and magnitude.

\begin{table}[H]
\centering
\caption{Summarizing the \ac{DNNs} performance for FABRIC using \ac{RMSE}.}
\label{tab:fabric_rmse}
\resizebox{\columnwidth}{!}{%
\begin{tabular}{cclll}
\hline
\textbf{Neural Network} & \textbf{Read Mean} & \textbf{Read StDev} & \textbf{Write Mean} & \textbf{Write StDev} \\ \hline
\textbf{FCN}           & 0.050              & 0.000               & 0.020               & 0.000               \\
\textbf{FCNPlus}      & 0.052              & 0.004               & 0.019               & 0.003               \\
\textbf{InceptionTime} & 0.055              & 0.005               & 0.025               & 0.010               \\
\textbf{InceptionTimePlus} & 0.053          & 0.005               & 0.034               & 0.016               \\
\textbf{OmniScaleCNN}  & 0.059              & 0.003               & 0.033               & 0.021               \\
\textbf{ResCNN}        & 0.054              & 0.005               & 0.024               & 0.007               \\
\textbf{ResNet}        & 0.051              & 0.003               & 0.066               & 0.038               \\
\textbf{ResNetPlus}    & 0.061              & 0.010               & 0.030               & 0.012               \\
\textbf{TCN}           & 0.050              & 0.000               & 0.021               & 0.003               \\
\textbf{XCM}           & 0.065              & 0.014               & 0.020               & 0.005               \\
\textbf{XCMPlus}       & 0.065              & 0.009               & 0.026               & 0.005               \\ \hline
\end{tabular}%
}
\end{table}

\begin{figure}[htbp]
    \centering
    \subfloat[\centering \scriptsize{FIBRE-NG: ResNet Training Process (Read).}]{
        \includegraphics[width=0.45\columnwidth]{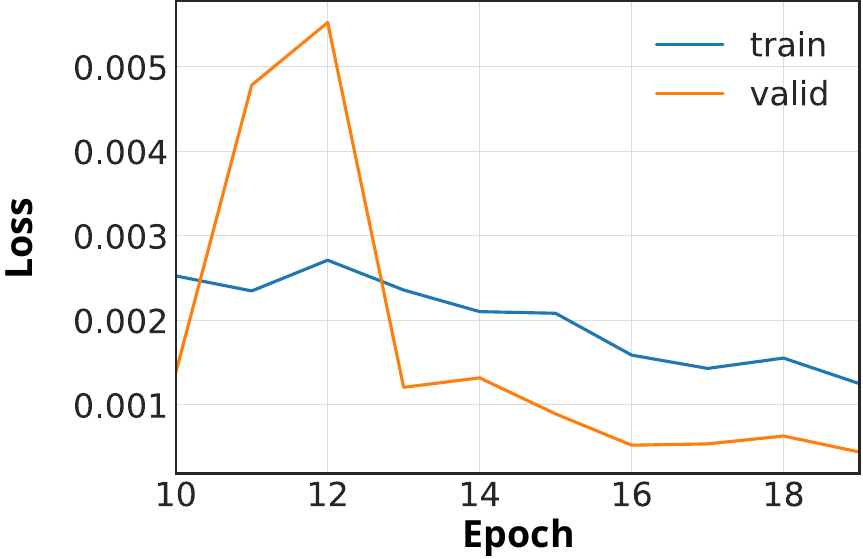}
    }
    \subfloat[\centering \scriptsize{FIBRE-NG: ResNet Training Process (Write).}]{
        \includegraphics[width=0.45\columnwidth]{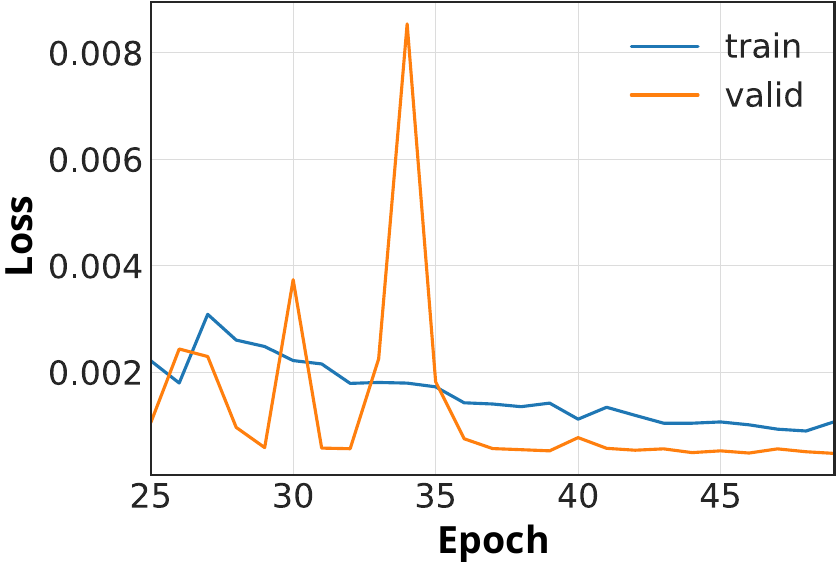}
    }
    \\
    \subfloat[\centering \scriptsize{Fabric: InceptionTimePlus Training Process (Read).}]{
        \includegraphics[width=0.45\columnwidth]{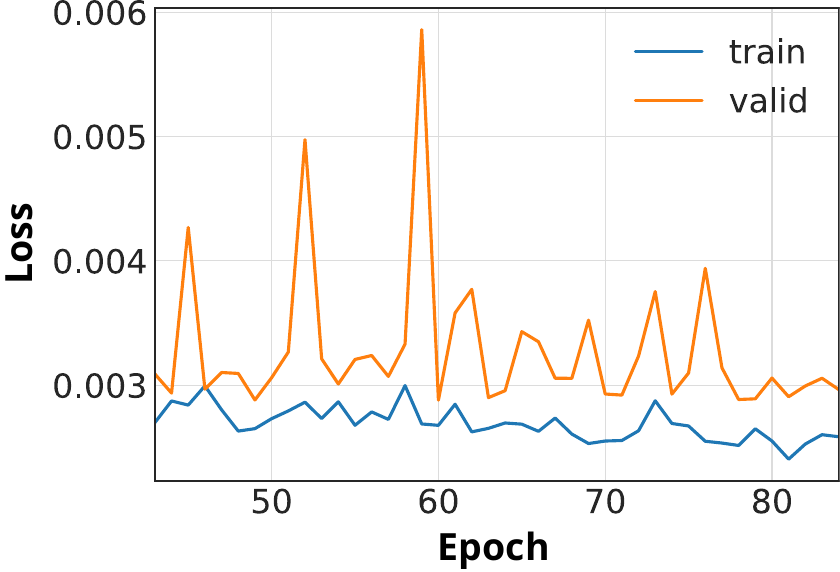}
    }
    \subfloat[\centering \scriptsize{Fabric: TCN Training Process (Write).}]{
        \includegraphics[width=0.45\columnwidth]{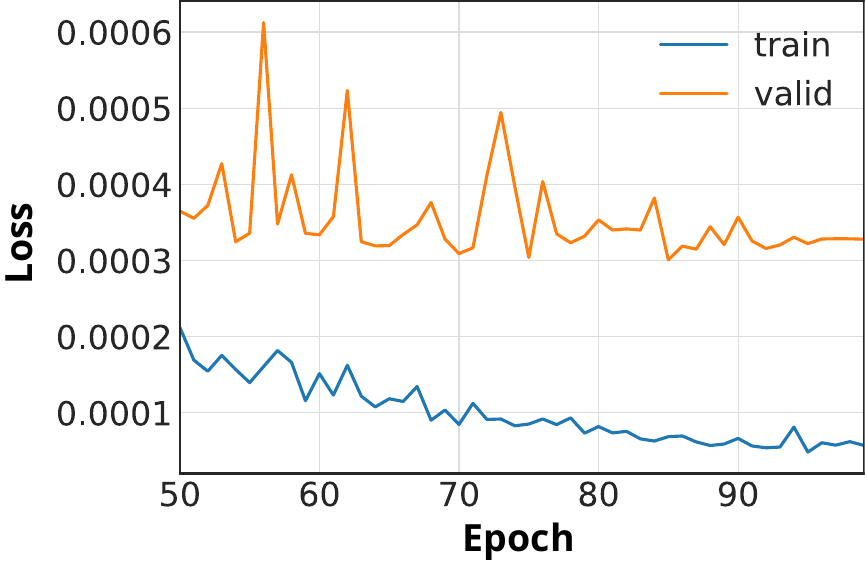}
    }
    \caption{\textcolor{black}{Training and Test behavior for Read and Write datasets on FIBRE-NG and Fabric Testbeds.}}
    \label{fig:training_behavior_all}
\end{figure}

For the read and write datasets from the \ac{FIBRE-NG} testbed shown in Fig.~\ref{fig:training_behavior_all}-a and Fig.~\ref{fig:training_behavior_all}-b, we observed the training behavior of ResNet \ac{DNN}. From Fig.~\ref{fig:training_behavior_all}-c, which refers to \ac{DNN} InceptionTimePlus for the read dataset in the Fabric testbed, there was a subtle drop in the training and test losses, indicating that \ac{DNNs} could extract patterns from the time series for prediction. As with Fig.~\ref{fig:training_behavior_all}-d, which refers to the dataset written in the Fabric testbed, there is a visual indication that the loss decreases as the epochs advance.

The results presented in Table~\ref{tab:fabric_rmse} and Table~\ref{tab:fibre_rmse} summarize the performance of \ac{DNNs} in the context of FIBRE-NG, utilizing the \ac{RMSE} metric. The InceptionTimePlus network stands out with the lowest RMSE for both reading (0.011) and writing (0.016), indicating superior performance compared to other networks. The ResNet also shows competitive results, with RMSE values of 0.010 for reading and 0.012 for writing. In contrast, the XCM and OmniScaleCNN exhibit the highest \ac{RMSE} values, suggesting lower accuracy in their predictions. Additionally, the table reveals the variability of the results, reflected in the standard deviations (StDev), which vary across the different architectures, with most maintaining relatively low deviations, especially in reading tasks.

\begin{table}[H]
\centering
\caption{Summarizing the \ac{DNNs} performance for FIBRE-NG using \ac{RMSE}.}
\label{tab:fibre_rmse}
\resizebox{\columnwidth}{!}{%
\begin{tabular}{cclll}
\hline
\textbf{Neural Network} & \textbf{Read Mean} & \textbf{Read StDev} & \textbf{Write Mean} & \textbf{Write StDev} \\ \hline
\textbf{FCN}           & 0.017              & 0.008               & 0.021               & 0.007               \\
\textbf{FCNPlus}      & 0.013              & 0.005               & 0.014               & 0.005               \\
\textbf{InceptionTime} & 0.013              & 0.005               & 0.024               & 0.027               \\
\textbf{InceptionTimePlus} & 0.011          & 0.003               & 0.016               & 0.007               \\
\textbf{OmniScaleCNN}  & 0.020              & 0.009               & 0.035               & 0.025               \\
\textbf{ResCNN}        & 0.021              & 0.019               & 0.026               & 0.013               \\
\textbf{ResNet}        & 0.010              & 0.000               & 0.012               & 0.004               \\
\textbf{ResNetPlus}    & 0.014              & 0.005               & 0.017               & 0.005               \\
\textbf{TCN}           & 0.019              & 0.007               & 0.015               & 0.005               \\
\textbf{XCM}           & 0.025              & 0.007               & 0.035               & 0.012               \\ \hline
\end{tabular}%
}
\end{table}

Having the \ac{DNNs} learn over the epochs operating on the datasets generated in the two testbeds, it is possible to admit that the \ac{DNNs} are efficient in dealing with the seasonality of a production-ready network. Then, slicing orchestrators can couple such models into their slicing management control loop and modernize the delivery of service verticals with a guaranteed \ac{SLA}.

\begin{figure}[H]
    \centering
    \subfloat[FIBRE-NG: Write Latency (ResNet)]{
        \includegraphics[width=0.45\columnwidth]{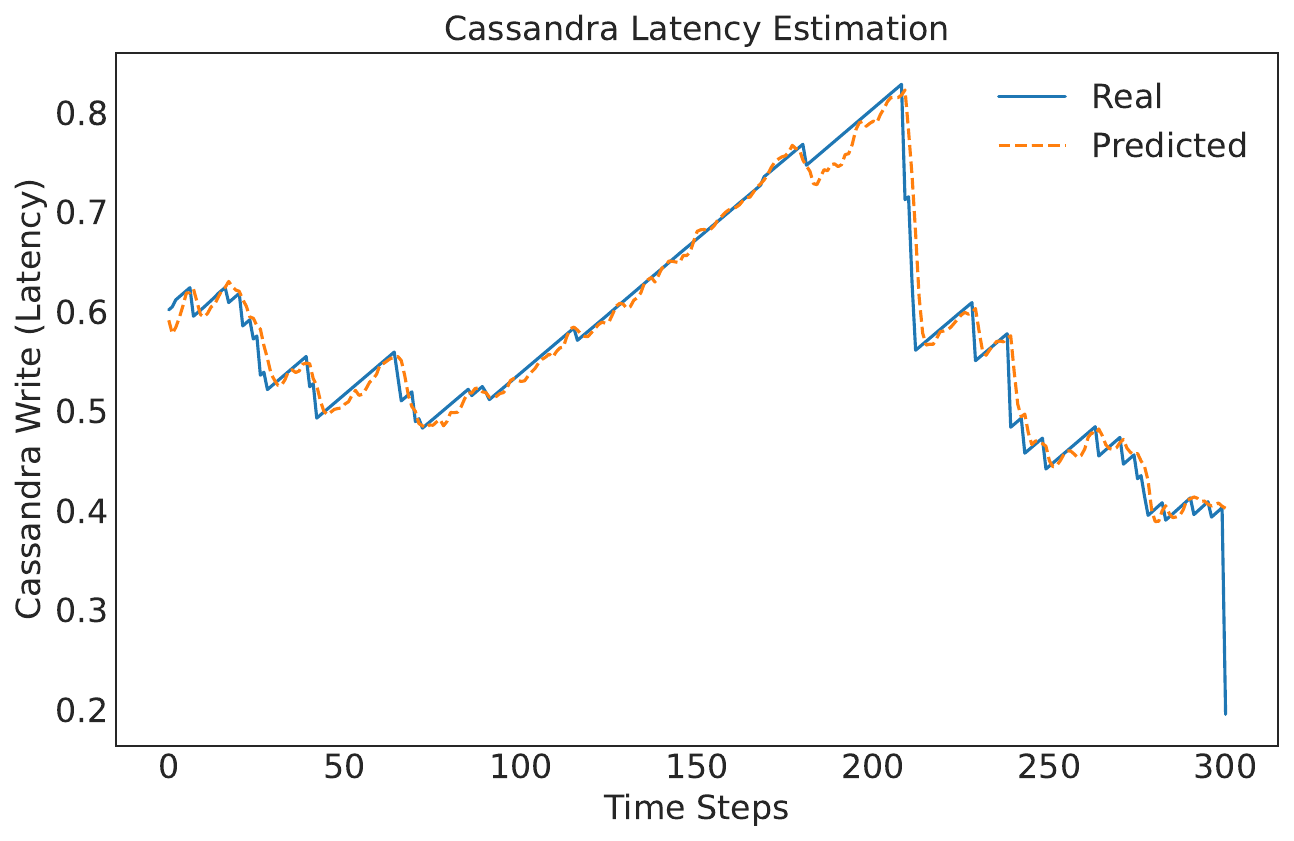}
    }
    \subfloat[FIBRE-NG: Read Latency (InceptionTimePlus)]{
        \includegraphics[width=0.45\columnwidth]{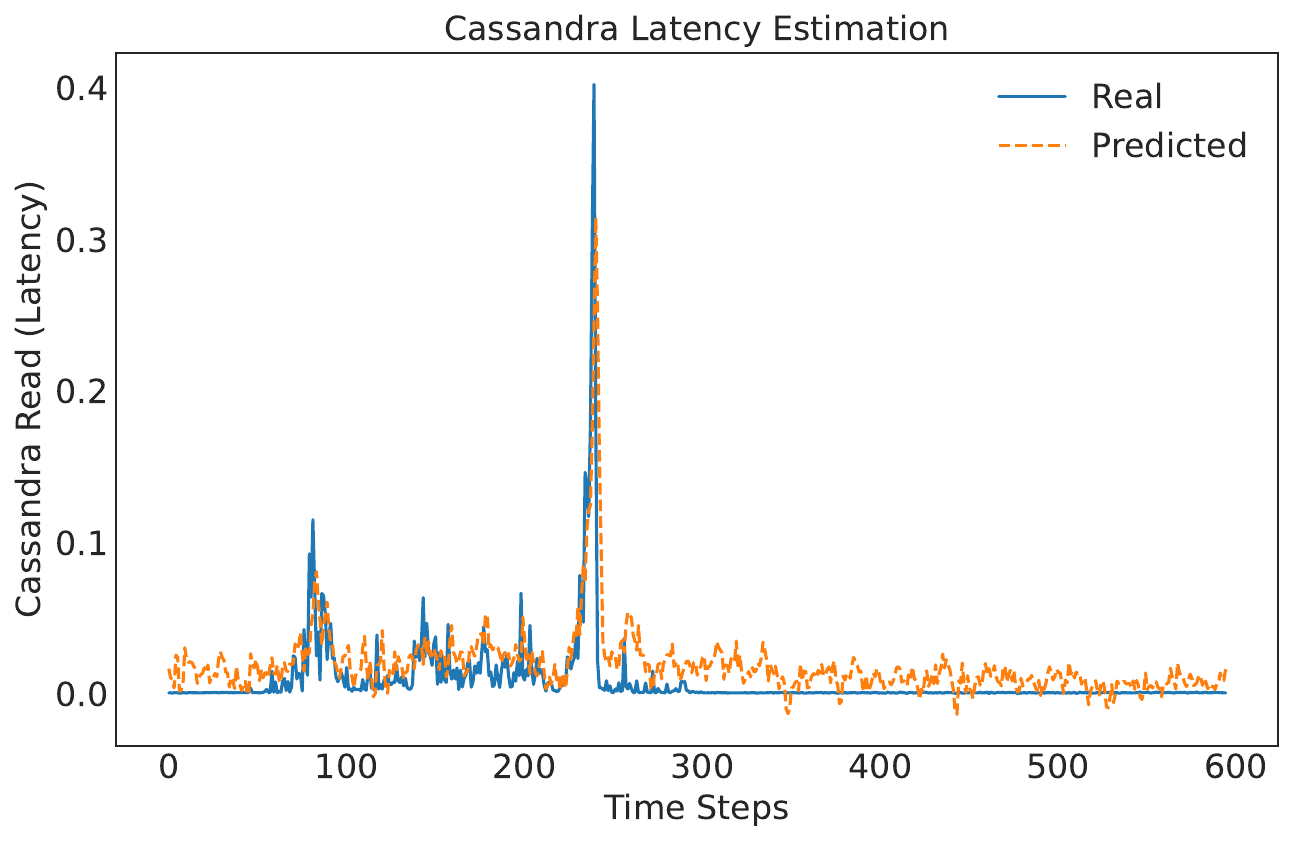}
    }
    \\
    \subfloat[Fabric: Write Latency (TCN)]{
        \includegraphics[width=0.45\columnwidth]{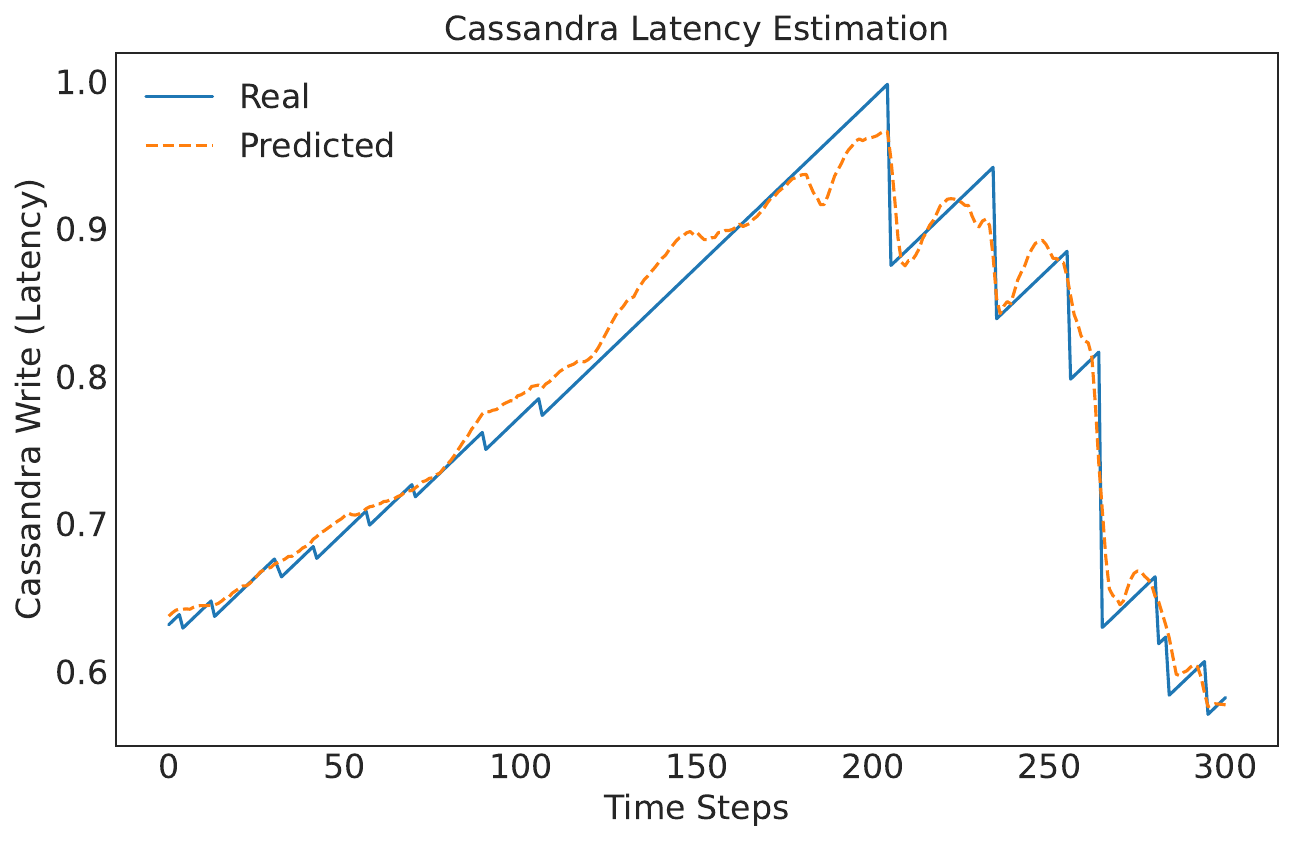}
    }
    \subfloat[Fabric: Read Latency (ResNet)]{
        \includegraphics[width=0.45\columnwidth]{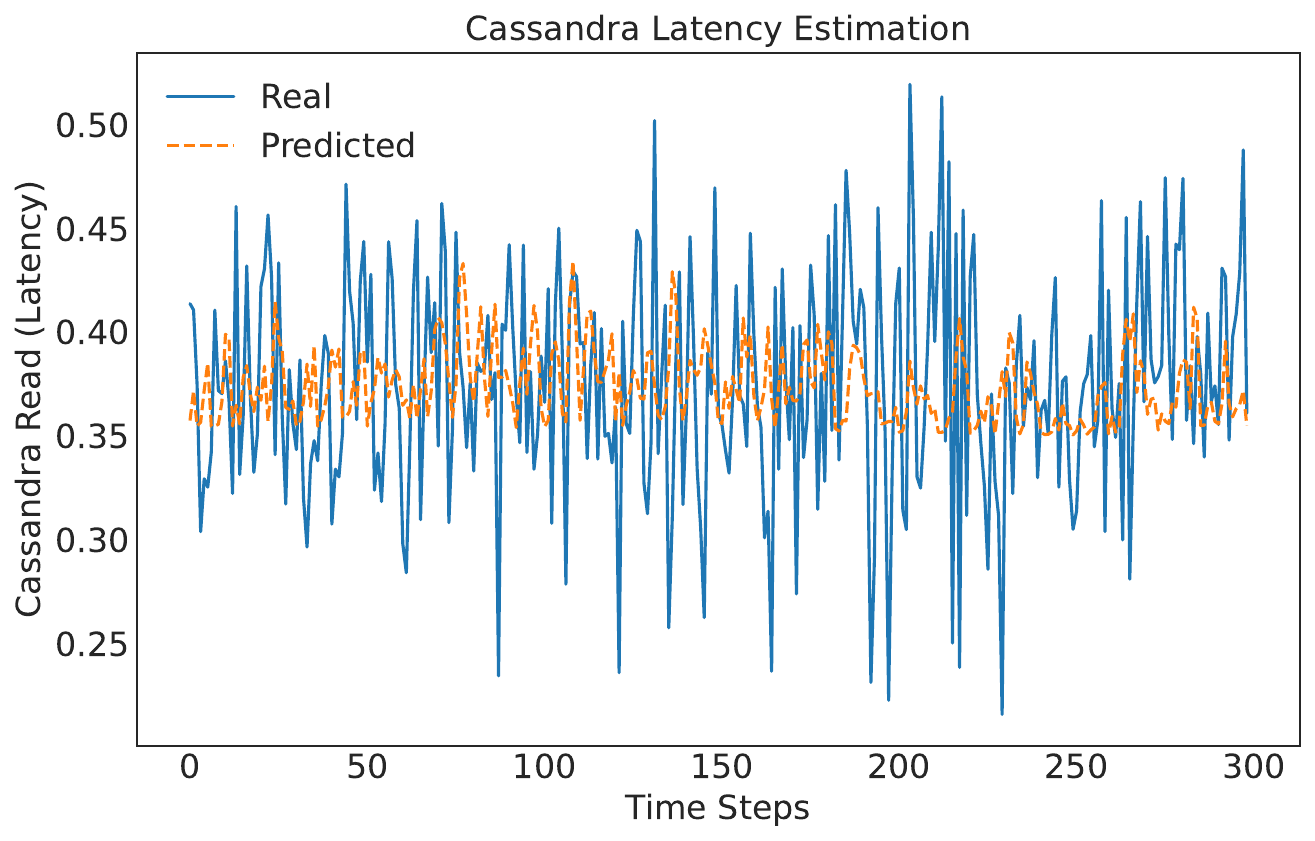}
    }
    \caption{\textcolor{black}{Latency prediction for better \ac{DNNs} for Write and Read operations in testbeds.}}
    \label{fig:real_predicted_by_better_CNNS}
\end{figure}

We seek to observe the behavior of \ac{DNNs} in predicting Cassandra latency by contrasting the real and predicted in the test portion of the time series.  Fig.~\ref{fig:real_predicted_by_better_CNNS} shows the \ac{DNNs} and their prediction process performance, considering the lowest value of the \ac{MAE} metric for both the Write or Read operations presented in Table~\ref{tab:fabric_mae} and Table~\ref{tab:fibre_mae}. Thus, modern slicing orchestrations can adopt a threshold for the difference between actual and predicted and assess whether the network slices it manages comply with the agreed \ac{SLA}.

\begin{figure}[H]
    \centering
    \subfloat[FIBRE-NG: MAPE (Write)]{
        \includegraphics[width=0.8\columnwidth]{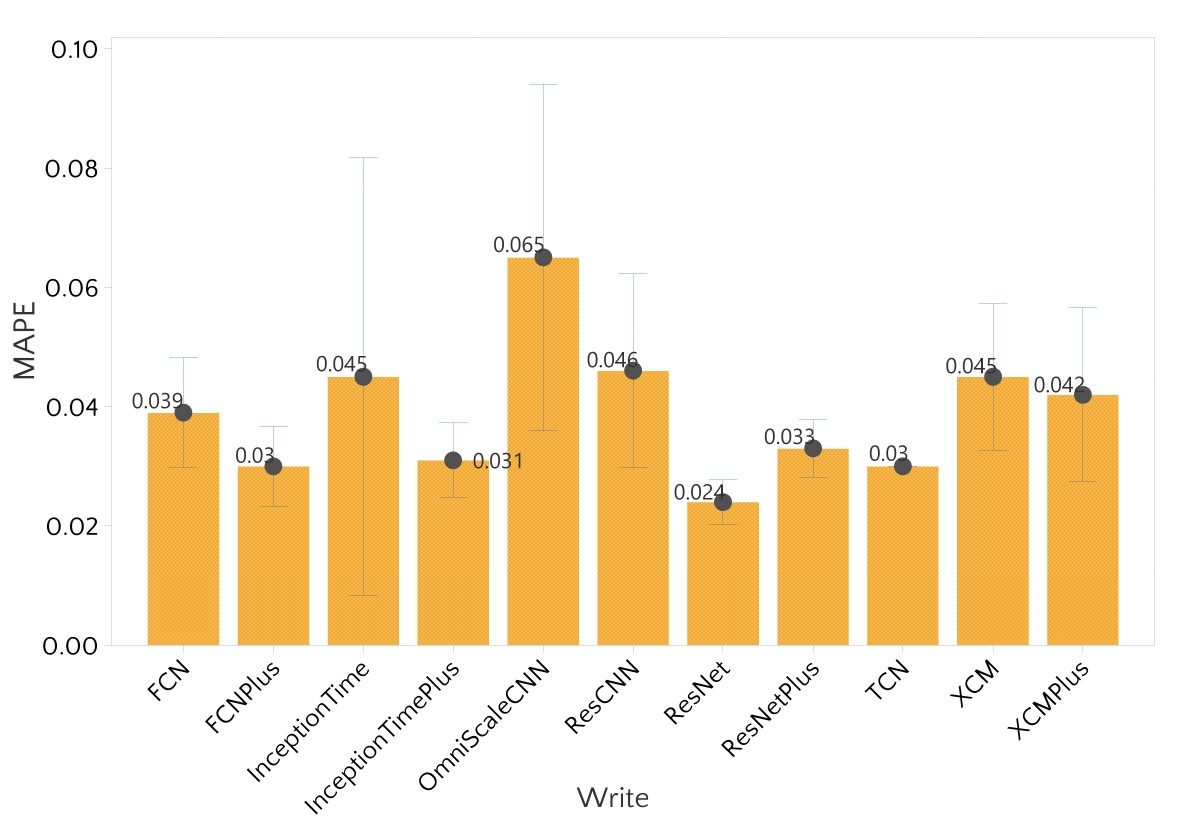}
    }
    \\
    \subfloat[FIBRE-NG: MAPE (Read)]{
        \includegraphics[width=0.8\columnwidth]{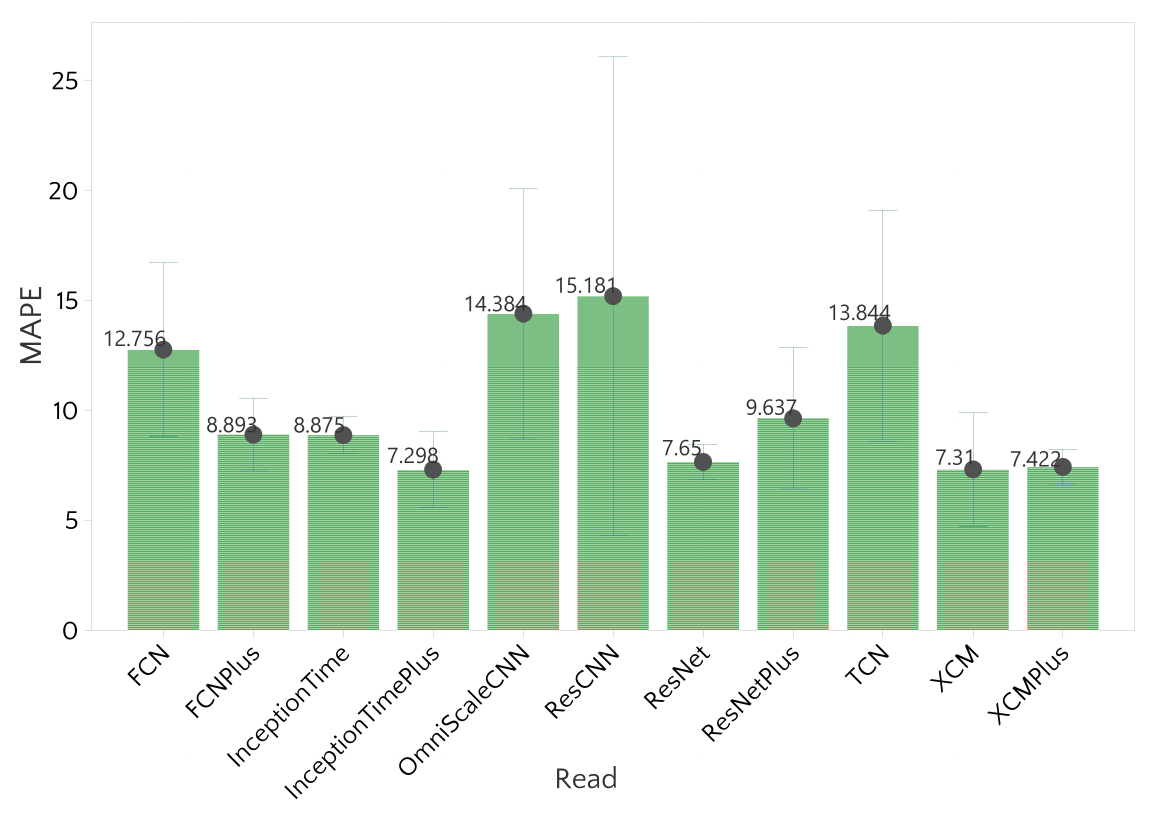}
    }
    \caption{Mean Absolute Percentage Error (MAPE) for FIBRE-NG in Write and Read operations.}
    \label{fig:FIBRE_MAPE}
\end{figure}

To further our analysis, we examined the performance of \ac{DNNs} for the four datasets generated in the two testbeds using the \ac{MAPE} metric. As illustrated in Fig.~\ref{fig:FIBRE_MAPE}-a, for the \ac{FIBRE-NG} testbed, ResNet demonstrated the highest performance on the Write dataset with a \ac{MAPE} of $0.024$. Conversely, for the Read dataset, \ac{DNN} InceptionTimePlus exhibited superior performance, as depicted in Fig.~\ref{fig:FIBRE_MAPE}-b.

\begin{figure}[!tbp]
    \centering
    \subfloat[Fabric: MAPE (Write)]{
        \includegraphics[width=0.8\columnwidth]{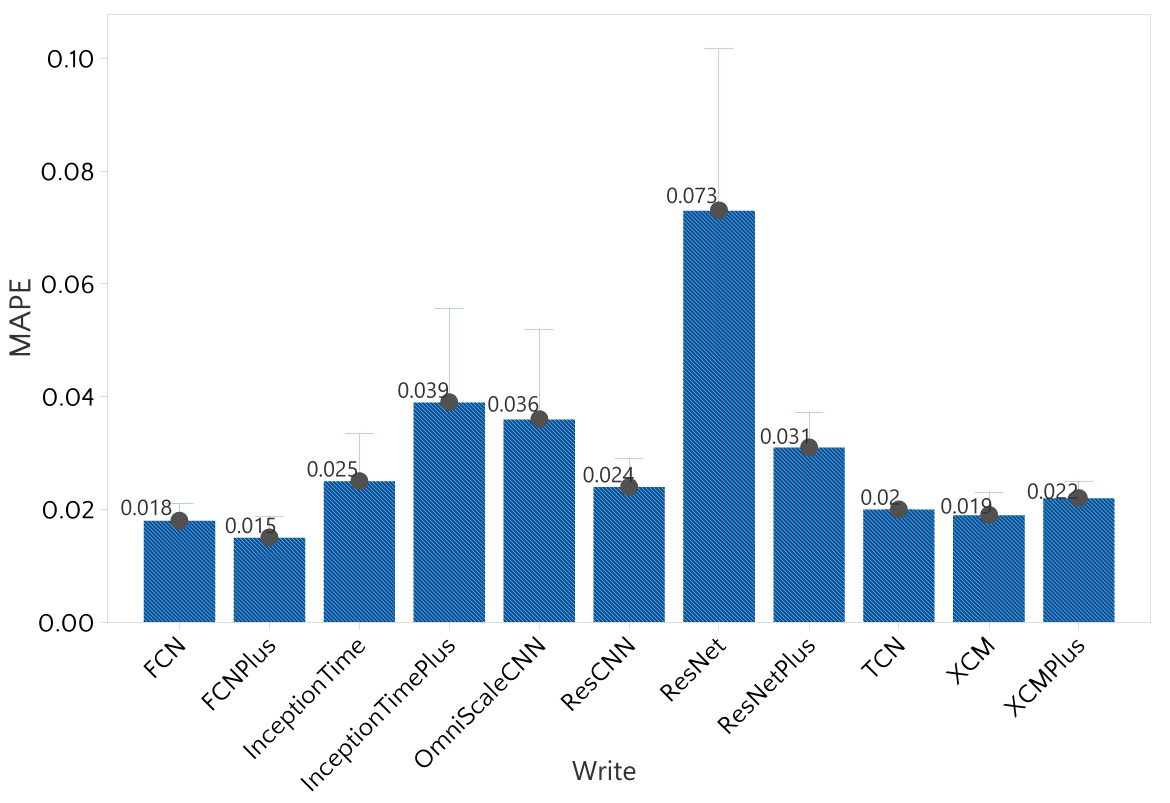}
    }
    \\
    \subfloat[Fabric: MAPE (Read)]{
        \includegraphics[width=0.8\columnwidth]{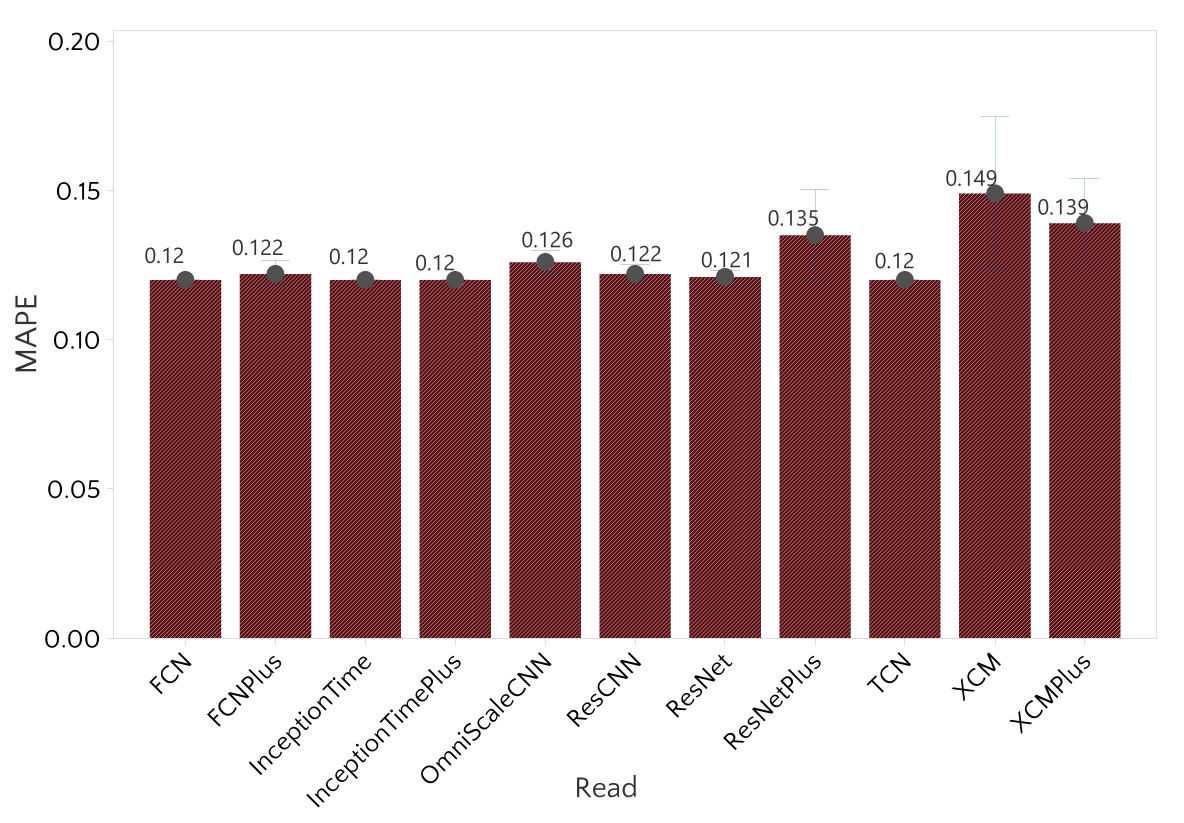}
    }
    \caption{Mean Absolute Percentage Error (MAPE) for Fabric in Write and Read operations.}
    \label{fig:FABRIC_MAPE}
\end{figure}

Furthermore, our analysis revealed that the \ac{DNN} exhibiting optimal performance for the Write dataset in the Fabric testbed was FCNPlus (Fig.~\ref{fig:FABRIC_MAPE}-a), achieving a \ac{MAPE} of $0.015$, with standard deviation serving as the criterion for resolving ties. 

In the Fabric Read scenario, and according to Fig.~\ref{fig:FABRIC_MAPE}-b, the \ac{MAPE} showed very low and homogeneous values across the models, indicating high prediction accuracy. The low variability suggests that the prediction task for this operation was relatively simple, resulting in insignificant percentage errors. Models such as FCN, ResNet, and TCN performed practically identically, whereas XCM and XCMPlus showed slight variations, but still within a very small error range.

Considering this empirical analysis of the behavior of \ac{DNNs} and the \ac{MAPE} metric, it is possible to admit that \ac{DNNs} are technologies that perform intelligent network slicing. It is possible to couple such capabilities into different building blocks to act at different phases of the lifecycle of a network slice. Using the \ac{MAPE} metric, it is possible to have a percentage dimension of the error of \ac{DNNs} that reiterates its ability to be embedded in prediction \ac{APIs} based on microservices, such as \ac{SFI2} Orchestration Architecture.

\textcolor{black}{Our approach primarily focused on evaluating a distributed database to assess the feasibility of training \ac{ML} algorithms on a large-scale testbed. In addition, our method can be extended to a broader range of applications that require access to computing and network-monitoring platforms. This extension would enable the integration of diverse performance metrics, facilitating accurate application performance forecasting and supporting life-cycle decision making in network slicing architectures.}

\section{Concluding Remarks}\label{sec:concluding_remarks}

In this study, we shed light on how \ac{ML} techniques, specifically \ac{DNNs} \textcolor{black}{and basic \ac{ML} algorithms}, can be jointly employed with slicing orchestration architectures to leverage and guarantee \ac{SLA} for tailored applications in nationwide testbeds. To achieve this, we propose a method for generating and aggregating datasets regarding the latency of \textbf{W}rite and \textbf{R}ead operations in a distributed database. We found that there are approaches in the literature that combine computational intelligence for the different phases of the network slice life cycle; however, they have not yet considered how these \ac{AI} techniques behave in production-ready networks deployed on nationwide testbeds.

Among the findings, we found that \ac{DNNs} \textcolor{black}{and even basic \ac{ML} algorithms} are promising technologies that can be built or natively embedded in slicing architectural building blocks to perform zero-touch orchestration in production-ready networks. Furthermore, we verified that forecasting network slicing latency with a low error rate is possible by monitoring generic and easily collected metrics related to the computing or network resources on which the network slice is deployed. Furthermore, we believe that embedding the \ac{DNN} or \textcolor{black}{basic \ac{ML}} models in \ac{SFI2} \ac{AI} management to cope with stringent application vertical requirements is a promising path.

One of the constraints of this study is that it focuses on generic networks and computing metrics. We aimed to incorporate more heterogeneous metrics into the dataset construction process to assess the generalization of these metrics and achieve low error rates in our predictions. Currently, we are working on analyzing the separation of metrics to validate the impact of each on the final ability to estimate \ac{SLA} conformance and the employment of Reinforcement Learning. 

In addition, we plan to explore methods \textcolor{black}{ such as queueing theory, extreme value analysis, or bursty traffic models to better capture extreme network conditions using supervised learning methods} and other \ac{DNNs} \textcolor{black}{and attention-based mechanisms } to determine their efficacy in such contexts. Our results offer valuable insights and opportunities for the further exploration of intelligent native slicing architectures.


\section*{Declaration of Competing Interest}

The authors declare that they have no known competing financial interests or personal relationships that could have appeared to
influence the work reported in this paper.

\section*{Acknowledgment}

We acknowledge the financial support of the Brazilian National Council for Scientific and Technological Development (CNPq), grant \#421944/2021-8, and the FAPESP MCTIC/CGI Research project 2018/23097-3 - SFI2 - Slicing Future Internet Infrastructures. The authors also thank CNPq, CAPES, FAPESP and Instituto ANIMA.





\bibliographystyle{elsarticle-num} 
\bibliography{references}



\end{document}